\documentclass[12pt,preprint]{aastex61}

\begin{document}

\title{Contact binaries at the short period cut-off\\ I. statistics and the first photometric investigations of ten totally eclipsing systems}

\correspondingauthor{Li, Kai}
\email{kaili@sdu.edu.cn }

\author{Li, Kai}
\affiliation{Shandong Provincial Key Laboratory of Optical Astronomy and Solar-Terrestrial Environment, Institute of Space Sciences, Shandong University, Weihai, 264209, China}
\affiliation{Key Laboratory for the Structure and Evolution of Celestial Objects, Chinese Academy of Sciences}

\author{Xia, Qi-Qi}
\affiliation{Shandong Provincial Key Laboratory of Optical Astronomy and Solar-Terrestrial Environment, Institute of Space Sciences, Shandong University, Weihai, 264209, China}
\author{Michel, Raul}
\affiliation{Observatorio Astron\'{o}mico Nacional, Instituto de Astronom\'{\i}a, Universidad Nacional Aut\'{o}noma de M\'{e}xico, Apartado Postal 877, Ensenada, B.C. 22830, M\'{e}xico}
\author{Hu, Shao-Ming}
\affiliation{Shandong Provincial Key Laboratory of Optical Astronomy and Solar-Terrestrial Environment, Institute of Space Sciences, Shandong University, Weihai, 264209, China}
\author{Guo, Di-Fu}
\affiliation{Shandong Provincial Key Laboratory of Optical Astronomy and Solar-Terrestrial Environment, Institute of Space Sciences, Shandong University, Weihai, 264209, China}
\author{Gao, Xing}
\affiliation{Xinjiang Astronomical Observatory, 150 Science 1-Street, Urumqi 830011, China}
\author{ Chen, Xu}
\affiliation{Shandong Provincial Key Laboratory of Optical Astronomy and Solar-Terrestrial Environment, Institute of Space Sciences, Shandong University, Weihai, 264209, China}
\author{Gao, Dong-Yang}
\affiliation{Shandong Provincial Key Laboratory of Optical Astronomy and Solar-Terrestrial Environment, Institute of Space Sciences, Shandong University, Weihai, 264209, China}

\begin{abstract}
The period distribution of contact binaries exhibits a very sharp short period cut-off at 0.22 days. In order to provide valuable information on this short period limit, we observed ten totally eclipsing contact binaries with orbital periods near this cut-off. By detailed analysis using W-D code, we determined that two of these systems are A-subtype contact binaries while the others are W-subtype contact binaries and all the targets show shallow contact configurations. Half of the targets exhibit stellar spot activity and four of them have third light. A statistical work on well studied USPCBs was carried out, and physical parameters of fifty-five USPCBs were obtained. Some common properties for these systems were derived. Due to the study of the period-color diagram of USPCBs, we found that the period-color relation of USPCBs is different from other W UMa type contact binaries and USPCBs are metal poor old stars. In addition, the evolutionary states of these systems were discussed by constructing the color-density diagram. We derived that the evolutionary states of the two components of USPCBs show identical characteristics of other contact binaries despite slower evolutionary status caused by smaller mass.
We suggested that both the fully convective limit claimed by Rucinski (1992) and the AML theory recommended by Stepien (2006) can produce the short period cut-off and a tertiary companion is very important during the formation of the short period contact binaries by removing angular momentum from the host eclipsing pairs under certain circumstances.

\end{abstract}


\section{Introduction}
Contact binaries are generally composed of two late type stars that share a common convective envelope. The nearly equal minima reveal very close temperatures of the two components despite very different masses. This type of systems are proposed to be formed from short period detached binaries by angular momentum loss (AML) due to magnetic braking and tidal friction, ultimately the two components will merge into a single rapid rotation star or blue straggler (e.g., Bradstreet \& Guinan 1994; Eggleton \& Kisseleva-Eggleton 2002; Qian et al. 2006). A tertiary companion may play an essential role during the process by removing angular momentum and energy exchanges (e.g., Eggleton \& Kisseleva-Eggleton 2006; Fabrycky \& Tremaine 2007, Qian et al. 2014, 2018). From the observations of spectroscopic binaries, Tokovinin et al. (2006) discovered that most of the short period close binaries (P$<3$ days) are in multiple systems. Recently, the statistics carried out by Liao \& Qian (2010) suggested that contact binaries are more frequently existing in triple systems. Therefore, tertiary companions are indeed essential to understand the formation and evolution of contact binaries.

According to the classification by Binnendijk (1970), contact binaries can be divided into two groups: A-subtype and W-subtype. For A-subtype star, the hotter component is the more massive one and the primary minimum is originated by the transit of the cooler and less massive component. For W-subtype star, the less massive component is the hotter one and the primary minimum corresponds to its occultation. The debate on the evolutionary correlation between the two subtype contact binaries has been active for a long time. Recently, Yildiz \& Do\v{g}an (2013) found that the initial evolutionary parameters of the two groups are totally dissimilar.

The light curves of contact binaries usually show different brightness of the two maxima, and some of them exhibit variations in the light curves from season to season, such as AD Cnc (Qian et al. 2007a); EQ Tau (Li et al. 2014); KIC 9532219 (Lee et al. 2016), this is referred as the O'Connell effect (O'Connell 1951). Star-spots due to the active component(s) will result in this effect. Continuous photometric and spectroscopic observations of such systems are essential for their complete understanding.

The orbital period distribution of contact binaries shows a very sharp cut-off at 0.22 days (Rucinski 1992, 1997), and many researchers have tried to solve this issue. Rucinski (1992) theorized that the components of the contact binaries at the short period cut-off will reach the fully convective limit and become dynamically unstable. Stepien (2006) suggested that the time scale of the AML is so long that even at the age of universe contact binaries at the short period limit should not be found. According to Jiang et al. (2012), the main reason for the short period limit might be due to the unstable mass transfer when the low mass initial primary component fills its inner Roche lobe. More recently, Qian et al. (2015a) proposed that circumbinary companions play a vital role in the origin and evolution of the short period contact binaries and suggested that all ultrashort period M-dwarf binaries are possibly triple systems. At present, the explanation of the observed abrupt short period limit is still an open question. In order to comprehend this issue, more and more such contact binaries should be observed. Thanks to the worldwide surveys, such as Catalina Sky Surveys (CSS, Drake et al. 2014b), LINEAR (Palaversa et al. 2013), All Sky Automated Survey (ASAS, Pojmanski 1997; Pojmanski et al. 2005), and Northern Sky Variability Survey (NSVS, Wozniak 2004), a great number of contact binaries around the short period limit have been identified. This provides a chance to analyze their origin and evolution. Since 2015, we have observed many ultrashort period contact binaries (USPCBs) with periods below 0.23 days and carried out a series of works on photometric studies of such systems. Our research goal is to analyze the light curves of a large sample of contact binaries, with periods around the short period limit, and provide valuable information on the formation of contact binaries at the short period cut-off as well as the origin and evolution of contact binaries. In this first paper, we present the statistics of the orbital period distribution of contact binaries and the well studied contact binaries with periods less than 0.23 days, along with the first photometric investigations of the ten totally eclipsing systems shown in Table 1.

\begin{table*}
\tiny
\begin{center}
\caption{The information of the ten USPCBs}
\setlength{\leftskip}{-20pt}
\begin{tabular}{lcccccclc}
\hline
Star	                          & Short name    &	RA	        &  Dec	      	&Period (days) &HJD$_0$ &V (mag)&Amplitude&References                        \\
&&&&&2450000+&&&\\\hline
CRTS J003244.2+244707	          & J003244	      & 00 32 44.21 &	+24 47 07.8	  &0.225173	     &8017.1335(1)&  14.82	 &0.68     & (1)            \\
CRTS J020730.1+145623	          & J020730	      & 02 07 30.20 &	+14 56 22.9	  &0.223442	     &8109.1606(2)&  15.79	 &0.73     & (1)            \\
CRTS J034705.9+211309           & J034705       & 03 47 05.96 & +21 13 09.7   &0.229570      &  8481.0334(5)&  15.82   &0.53     & (1)            \\
1SWASP J050904.45-074144.4	    & J050904	      & 05 09 04.46 &	-07 41 44.2	  &0.229575	     &7388.1801(1)&  13.29	 &0.64     & (2), (3)       \\
CRTS J053317.3+014049	          & J053317	      & 05 33 17.21 &	+01 40 49.8	  &0.215652	     &7778.0287(2)&  14.88	 &0.52     & (1)            \\
CRTS J060855.6+622713	          & J060855	      & 06 08 55.61 &	+62 27 13.7	  &0.229320	     &7763.0185(2)&  15.49	 &0.68     & (1)            \\
1SWASP J151144.56+165426.4	    & J151144	      & 15 11 44.56 &	+16 54 26.4	  &0.219865	     &7507.1028(1)&  14.65	 &0.62     & (1), (3)       \\
CRTS J151631.0+382626	          & J151631	      & 15 16 31.08 &	+38 26 25.9	  &0.227192	     &8213.9352(1)&  16.60	 &0.71     & (1)            \\
1SWASP J220235.74+311909.7	    & J220235	     & 22 02 35.75 &	+31 19 09.5	  &0.220477	     &8022.1242(1)&  14.02	 &0.61$^a$ & (3)            \\
CRTS J232100.1+410736	          & J232100	     & 23 21 00.14 &	+41 07 36.8	  &0.211984	     &8016.1532(2)&  15.03	 &0.45     & (1)            \\
\hline
\end{tabular}
\end{center}
References: (1) Drake et al. 2014b; (2) Norton et al. 2011; (3) Lohr et al. 2013b;.\\
$^a$ The variability amplitude of this star is roughly determined by the public data of SuperWASP.\\

\end{table*}

\section{The still existing orbital period cut-off of contact binaries}
At present, a great deal of contact binaries have been discovered. The AAVSO International Variable Star Index (VSX, Watson 2006, Version 2018-05-28) has combined the contact binaries from individual discoveries in the Galaxy. We carried out a statistic analysis of contact binaries based on the VSX, GCVS (Samus et al. 2017), the Catalogue of Eclipsing Variables (CEV, Avvakumova et al. 2013) and the photometric surveys, CSS (Drake et al. 2014b), LINEAR (Palaversa et al. 2013), ASAS (Pojmanski 1997; Pojmanski et al. 2005), NSVS (Wozniak 2004), Trans-Atlantic Exoplanet Survey(TrES, Devor 2008), THU-NAOC Transient Survey (TNTS, Yao et al. 2015), SuperWASP (143 short period contact binaries, Lohr et al. 2013b), Kepler (Pr\v{s}a 2011, Version 2), the Optical Gravitational Lensing Experiment (OGLE, Soszy\'{n}ski et al. 2016), Siding Spring Survey(SSS, Drake et al. 2017). The distribution of the orbital period of contact binaries is displayed in Figure 1. We can see a clearly sharp decline at 0.22 days. The total number of contact binaries below the short period cut-off is 1932, the fraction is only 1.11\% of all known contact systems. Though the number of discovered contact binaries below the 0.22 days cut-off is increasing, the short period cut-off remains. Therefore, the observations and researches on contact binaries around the short period limit are imperative.

\begin{figure}
\begin{center}
\includegraphics[angle=0,scale=0.8]{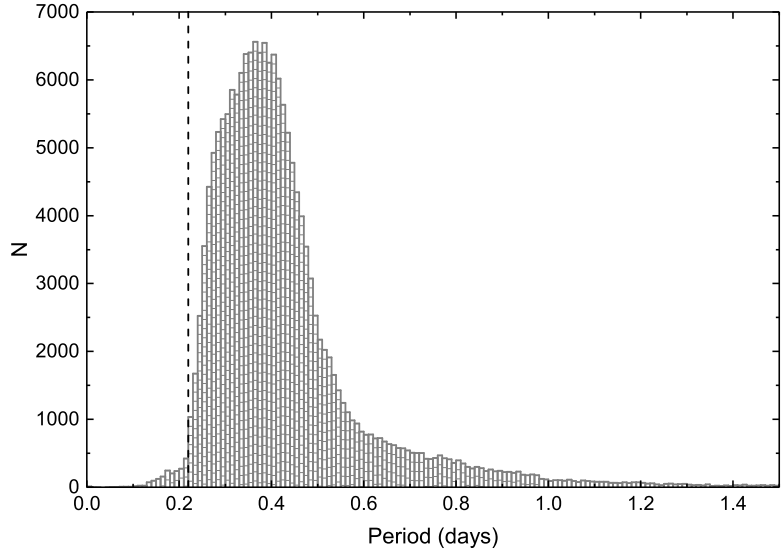}
\caption { The period distribution of contact binaries based on catalogs VSX, GCVS, CEV, CSS, LINEAR, ASAS, NSVS, TrES, TNTS, SuperWASP(143), Kepler, OGLE, SSS. The dashed line represents the sharp decline at 0.22 days.}
\end{center}
\end{figure}

\section{CCD observations of contact binaries at the short period cut-off }
In order to photometrically study the contact binaries at the short period cut-off, we have observed such systems by using the Weihai Observatory 1.0-m telescope of Shandong University (WHOT, Hu et al. 2014), the 2.12m telescope at the San Pedro Martir Observatory (SPMO2.12m), and the 85 cm telescope at the Xinglong Station of National Astronomical Observatories (NAOs85cm) in China. The CCD detectors are Andor DZ936 for WHOT,  E2V CCD42-40 for SPMO2.12m, and Andor DZ936 for NAOs85cm.
The field of views of the three telescopes are $12^{'}\times12^{'}$, $6^{'}\times6^{'}$, and $32^{'}\times32^{'}$, respectively. From 2015 to 2018, the complete light curves of 10 systems with totally eclpises have been obtained. The standard Johnson/Cousins filters were used during our observations. Most of the stars were observed using the $R_cI_c$ filters, only for $VR_cI_c$ one target, J151631, $BVR_cI_c$ light curves were obtained. The exposure times were in the range of 15 to 120s depending on the brightnesses of the stars, the filters used and weather conditions. The observational log for all the binaries is listed in Table 2. The IRAF software\nolinebreak \footnote{IRAF is distributed by the National Optical Astronomy Observatories, which is operated by the Association of Universities for Research in Astronomy Inc., under contract to the National Science Foundation.} was employed to reduce the CCD images. After the bias and flat corrections, the differential magnitudes for all the binaries were derived with respect to nearby comparison stars. For J151631, the standard star calibration was applied to determine the standard visual magnitude. Light curves of two examples of the stars are displayed in Figure 2.

\begin{table*}
\scriptsize
\begin{center}
\caption{The observational log for all the binaries}
\setlength{\leftskip}{-25pt}
\begin{tabular}{lcccccclc}
\hline
Target     &Observing Date	    &Exposure time	                & Comparison Star         & Check Star              & Telescope \\\hline
J003244	   &2017 Sep 20	        &$R_c$70s $I_c$40s	            & 2MASS J00325024+2446032 & 2MASS J00322836+2446126 & WHOT      \\
J020730	   &2017 Dec 22	        &$R_c$80s $I_c$50s	            & 2MASS J02073642+1456579 & 2MASS J02072161+1455417 & WHOT      \\
J034705    &2018 Dec 28         &$R_c$100s $I_c$70s             & 2MASS J03471058+2114230 & 2MASS J03471537+2117464 & NAOs85cm \\
J050904	   &2015 Dec 31	        &$V$50s $R_c$30s $I_c$20s	    & 2MASS J05090960-0740010 & 2MASS J05085797-0738211 & WHOT        \\
J053317	   &2017 Jan 24	        &$R_c$120s $I_c$40s	            & 2MASS J05331395+0140364 & 2MASS J05331448+0142248 & WHOT        \\
J060855	   &2017 Jan 09	        &$R_c$90s $I_c$70s	            & 2MASS J06090729+6226550 & 2MASS J06085199+6227573 & WHOT      \\
J151144	   &2016 Mar 21	        &$R_c$80s $I_c$60s	            & 2MASS J15114875+1653343 & 2MASS J15112699+1653205 & WHOT      \\
           &2016 Apr 28	        &$R_c$80s $I_c$60s	            & 2MASS J15114875+1653343 & 2MASS J15112699+1653205 & WHOT      \\
J151631$^a$	   &2018 Apr 05, 07, 09	&$B$40s $V$20s $R_c$15s $I_c$15s&  $-$                    &  $-$                & SPMO2.12m      \\
J220235	   &2017 Sep 25	        &$R_c$90s $I_c$60s	            & 2MASS J22023046+3117386 & 2MASS J22022651+3120238 & WHOT      \\
J232100	   &2017 Sep 19	        &$R_c$70s $I_c$40s	            & 2MASS J23210244+4110117 & 2MASS J23211031+4104560 & WHOT      \\   \hline
\end{tabular}
\end{center}
$^a$ For J151631, the standard star calibration was applied to determine the standard visual magnitude.
\end{table*}

\begin{figure*}
\begin{center}
\includegraphics[width=0.5\textwidth]{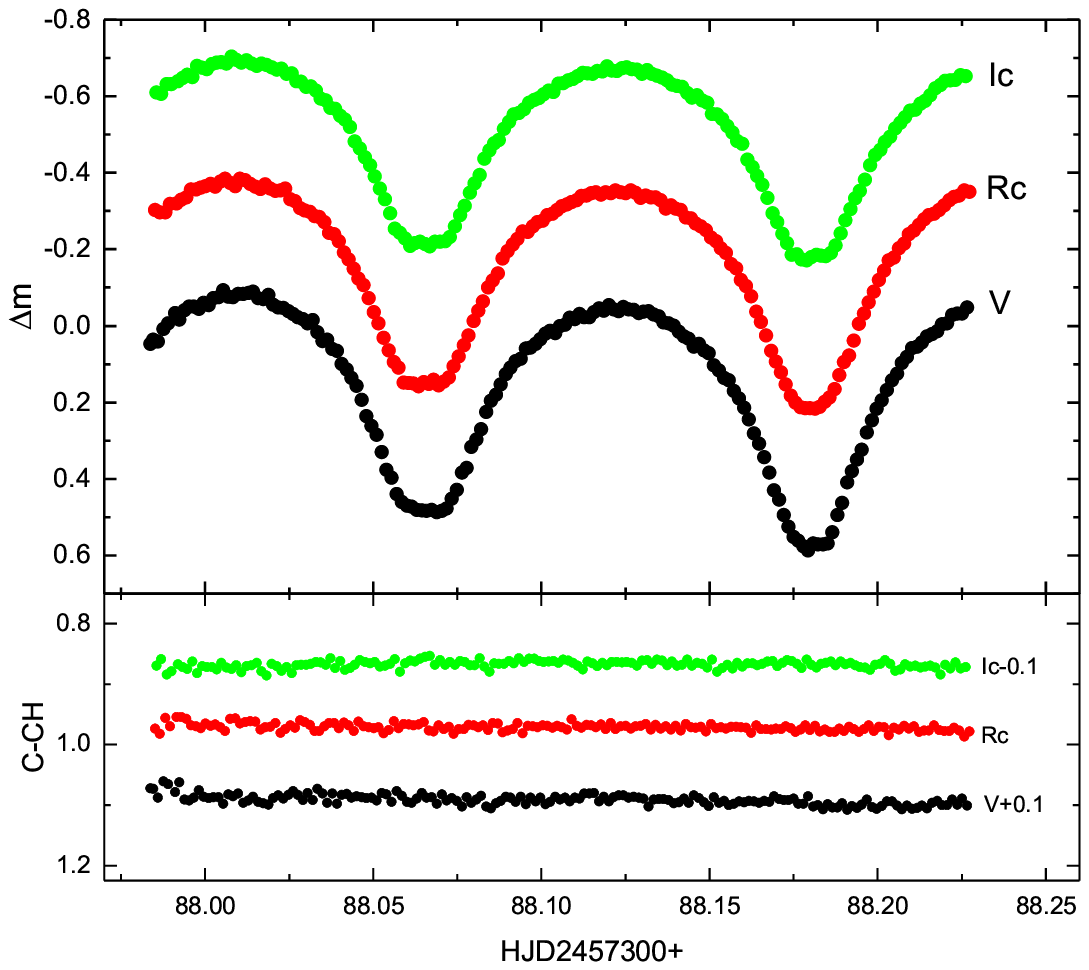}%
\includegraphics[width=0.51\textwidth]{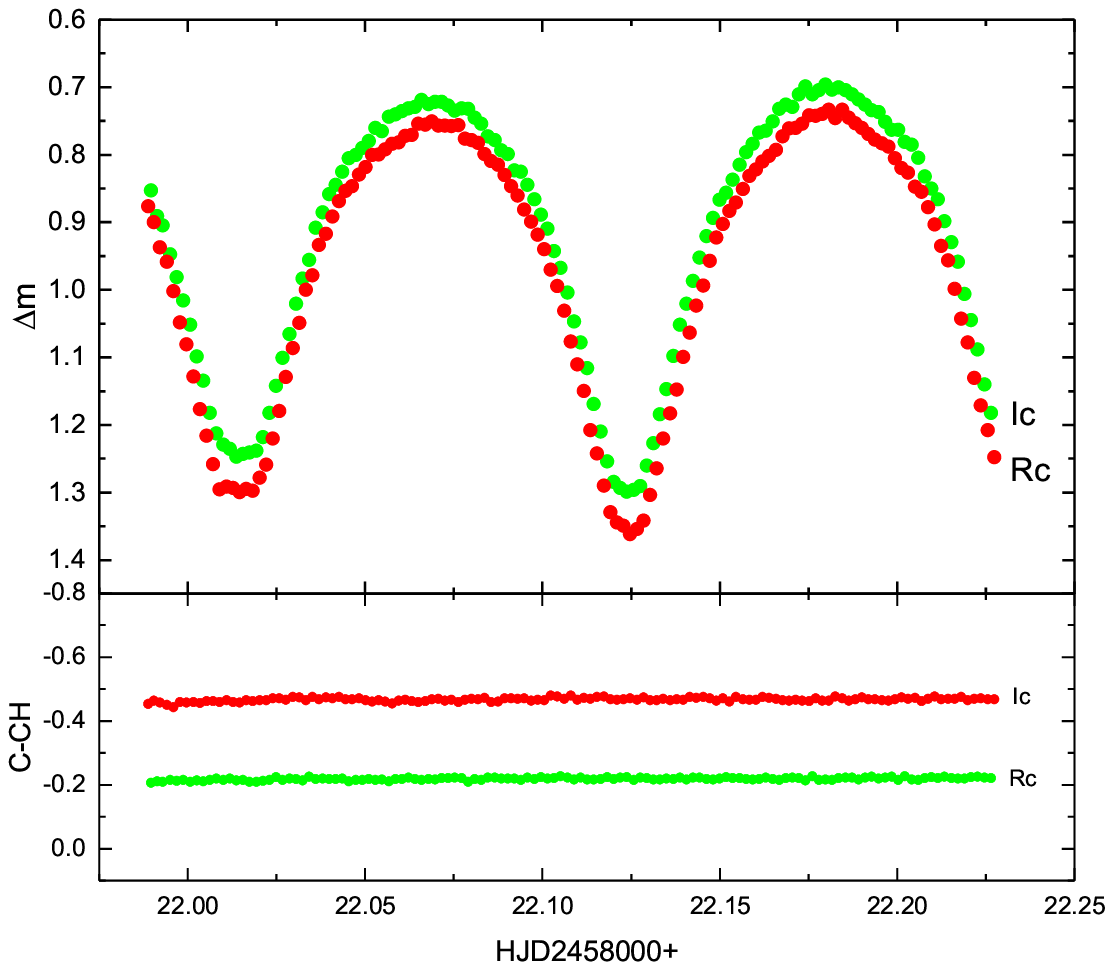}

\caption{The left panel plots the $VR_cI_c$ light curves of J050904 observed on December 31, 2015, while the right one displays the $R_cI_c$ light curves of J220235 observed on September 25, 2017.}
\end{center}
\end{figure*}

\section{Analyzing the light curves}
The Wilson-Devinney (W-D) code (Wilson \& Devinney 1971; Wilson 1979, 1990, 1994) was applied to model the light curves of all targets. In order to determine the effective temperatures of the targets, we did a bibliographical search and found that two of the targets have been observed by the Guoshoujing Telescope (the Large Sky Area Multi-Object Fiber Spectroscopic Telescope (LAMOST)) (Luo et al. 2015), one has been observed by  Koen et al. (2016), and all the targets have been observed by Gaia DR 2 (Gaia Collaboration et al. 2016, 2018), their effective temperatures, based on their spectroscopic observations, are listed in Table 3, if one star has two observations, the adopted effective temperature corresponds to the average value. These adopted effective temperatures are some sort of weighted average of the temperatures of both stars in the binary, not that of one star or the other, they are mean temperatures of the systems. During the modeling, the primary temperature was set as $T_1=T_m$. The bolometric and bandpass limb-darkening coefficients can be derived from van Hamme (1993) based on their temperatures. The gravity darkening and bolometric albedo coefficients of all targets were assumed to be $g_{1,2}=0.32$ and $A_{1,2} = 0.5$  because the temperatures of all systems are less than 6700 K, indicating that all the stars have convective envelopes.

During the modelling, we used mode 2 (detached mode) at first, later we found that the solutions are convergent at mode 3 (contact mode) for all systems. Since all the targets were analyzed for the first time, the mass ratios ($q$) were determined by using the $q-$search method by finding a series of solutions with assumed values of mass ratio. The relationships between the resulting sum of the squares of the residuals and the mass ratio for all the systems are shown in Figure 3. During the modelling, some parameters are adjustable, such as the orbital inclination, $i$, the effective temperature of star 2, $T_2$, the relative luminosity of star 1, $L_1$, and the dimensionless potential of star 1 $\Omega_1$. As shown in Figure 3, the minimum values of $\Sigma$ are derived for all targets. The corresponding mass ratio of the minimum value of $\Sigma$ for each system was then set as initial value and differential corrections were performed, the final physical parameters were derived when the solution converged. There are two exceptions, J050904 and J220235. For these two targets, we used the mass ratio of the minimum value of $\Sigma$ as an initial to determine the best physical parameters at first, the best fitting curves are displayed in the left panels of Figure 4 (the O-C residuals are the differences between the observed light curves and the fitting curves with hot spot and $L_3$), we can see that, according to the O-C residuals, it is not possible to obtain a good fit for both J050904 and J220235. Then, we used the mass ratio at the secondary minimum of the $q-$search figure as an initial value to determine the best physical parameters and the corresponding synthetic light curves are shown in the right panels of Figure 4. Very good agreement with the observed light curves can be seen. Therefore, we adopted the results derived by setting the mass ratio at the secondary minimum of the $q-$search figure as an initial value to be the final results for the two targets. The physical parameters of all targets are shown in Table 4, while the corresponding theoretical light curves  for the rest eight targets are displayed as black solid lines in Figure 5. Some of the targets show light curve asymmetries with the magnitudes at phase 0.25 being different from those at phase 0.75. This is called the O'Connell effect which is assumed to be caused by spot activity. In order to model the asymmetric light curves, the spots model of the W-D was used. Cool spot or hot spot on either of the two components was applied. In this process, the parameters of latitude ($\theta$), longitude($\lambda$), angular radius ($r_s$), and temperature factor ($T_s$) used to describe the spot were adjustable. The spot solutions are also listed in Table 4, and the synthetic light curves are shown as blue solid lines in Figure 5. Then, we considered to search third light for all targets and found that four of the systems have $L_3$. The synthetic light curves with third light are plotted using red solid lines in Figure 5. The solutions with $L_3$ and with or without spot are listed in Table 4 and chosen as the final results for the four systems. We should declare that the parameters of the spot were set as fixed, during third light search, in order to quickly determine a convergent solution. The final $T_1$ and $T_2$ were determined by the following equations (Coughlin et al. 2011; Kjurkchieva et al. 2018a),
\begin{eqnarray}
T_1&=&T_m+{c\Delta T\over c+1},  \\\nonumber
T_2&=&T_1-\Delta T,
\end{eqnarray}
where $\Delta T=T_1-T_2$ and $c=L_2/L_1$ are calculated based on the final photometric results. The final results of $T_1$ and $T_2$ are listed in Table 5.

\begin{table}
\begin{center}
\caption{The temperatures of the ten targets determined by Gaia DR 2, LAMOST and Koen et al. (2016).}
\begin{tabular}{lccccc}
\hline
Target   &  Gaia (K) &  LAMOST (K)  & Koen (K)      & $T_m$ (K) \\
\hline
J003244  &  4853     &  -           &  -           & 4853     \\
J020730  &  4311     &  -           &  -           & 4311     \\
J034706$^a$  &  4782     &  4899        &  -           & 4841     \\
J050904  &  4876     &  -           &  5340        & 5108     \\
J053317  &  4308     &  -           &  -           & 4308     \\
J060855  &  4265     &  -           &  -           & 4265     \\
J151144  &  4448     &  4405        &  -           & 4426     \\
J151631  &  4825     &  -           &  -           & 4825     \\
J220235  &  5001     &  -           &  -           & 5001     \\
J232100  &  4321     &  -           &  -           & 4321     \\
\hline
\end{tabular}
\end{center}
$^a$ the LAMOST effective temperature is an average value for this target.
\end{table}

\begin{figure}[hpb]\centering
\includegraphics[width=0.35\textwidth]{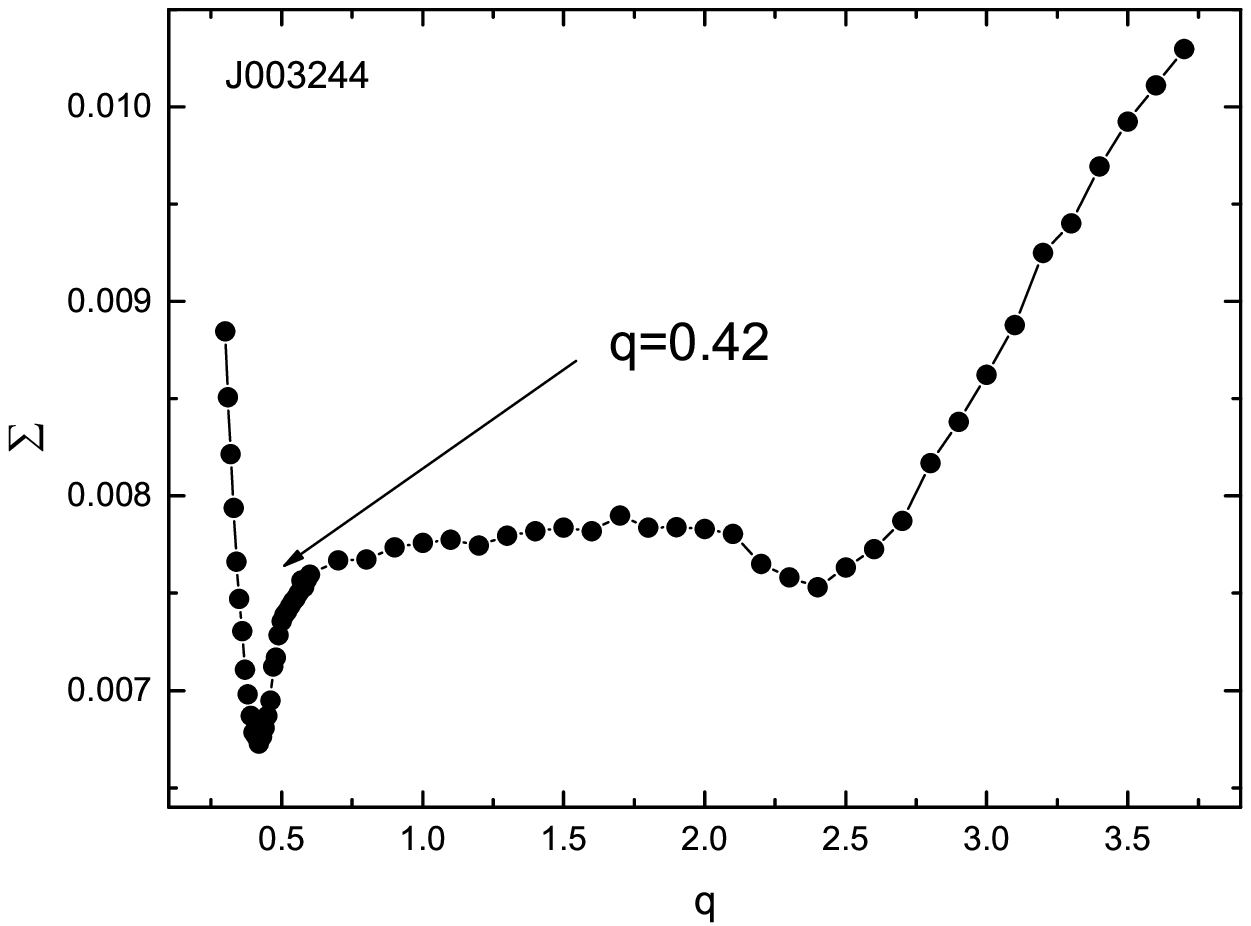}
\includegraphics[width=0.35\textwidth]{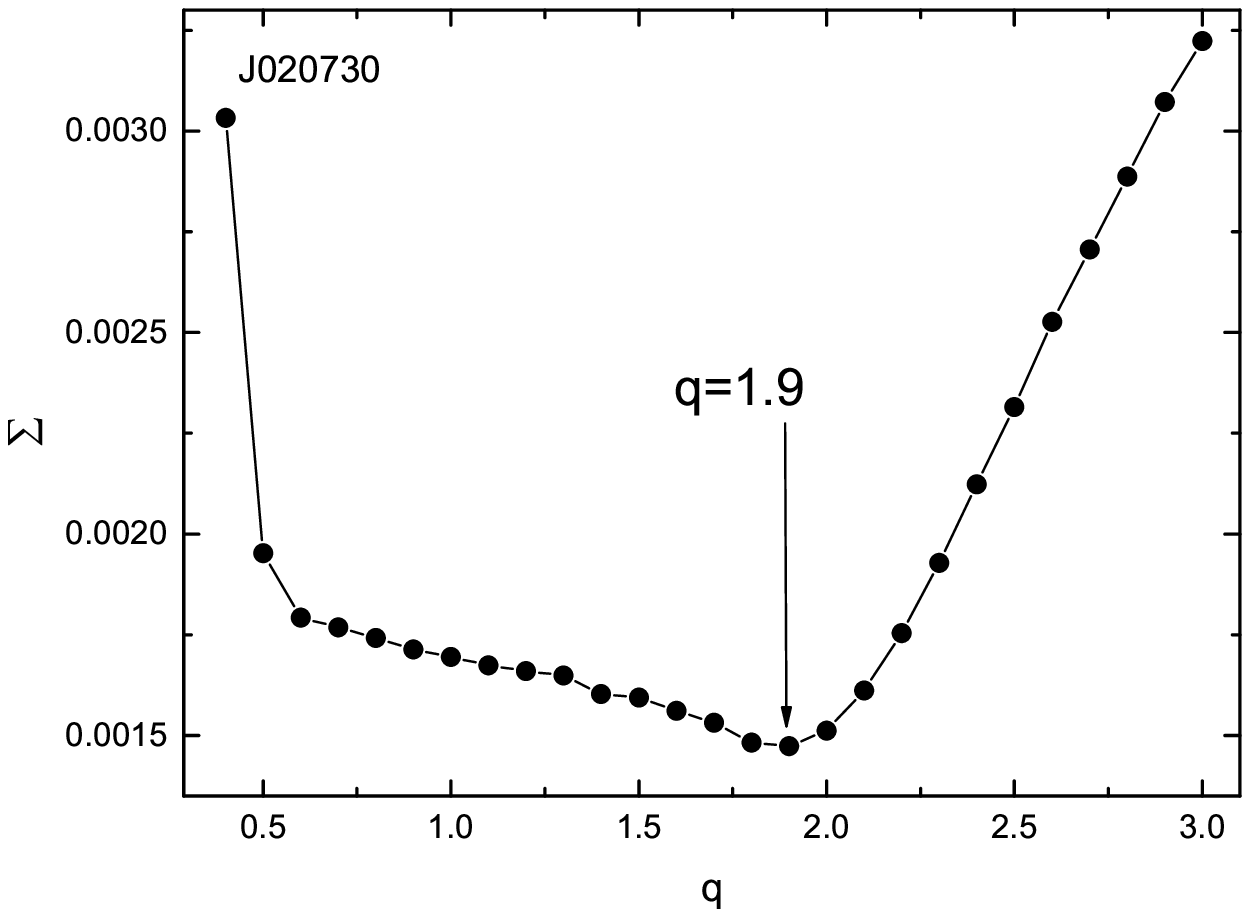}
\includegraphics[width=0.35\textwidth]{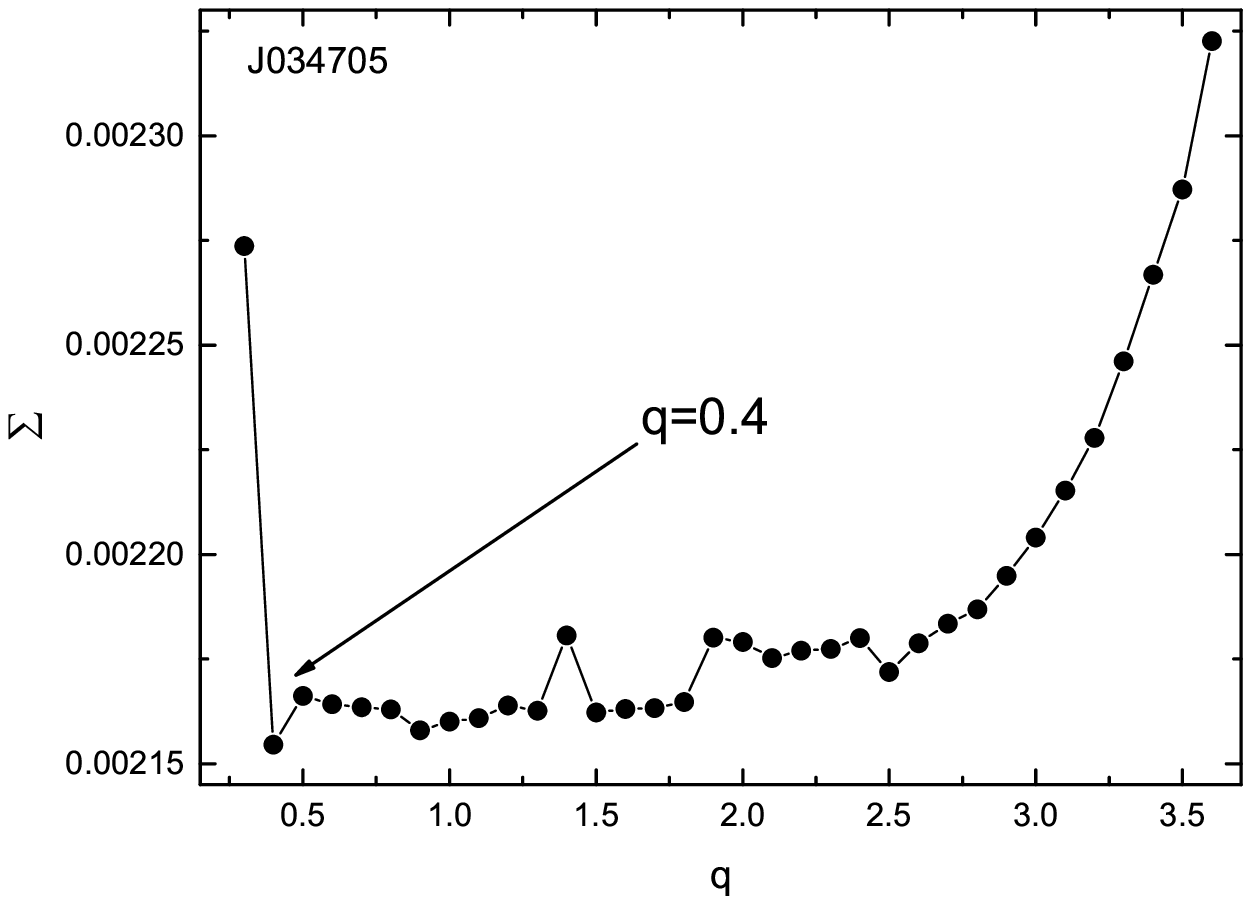}
\includegraphics[width=0.35\textwidth]{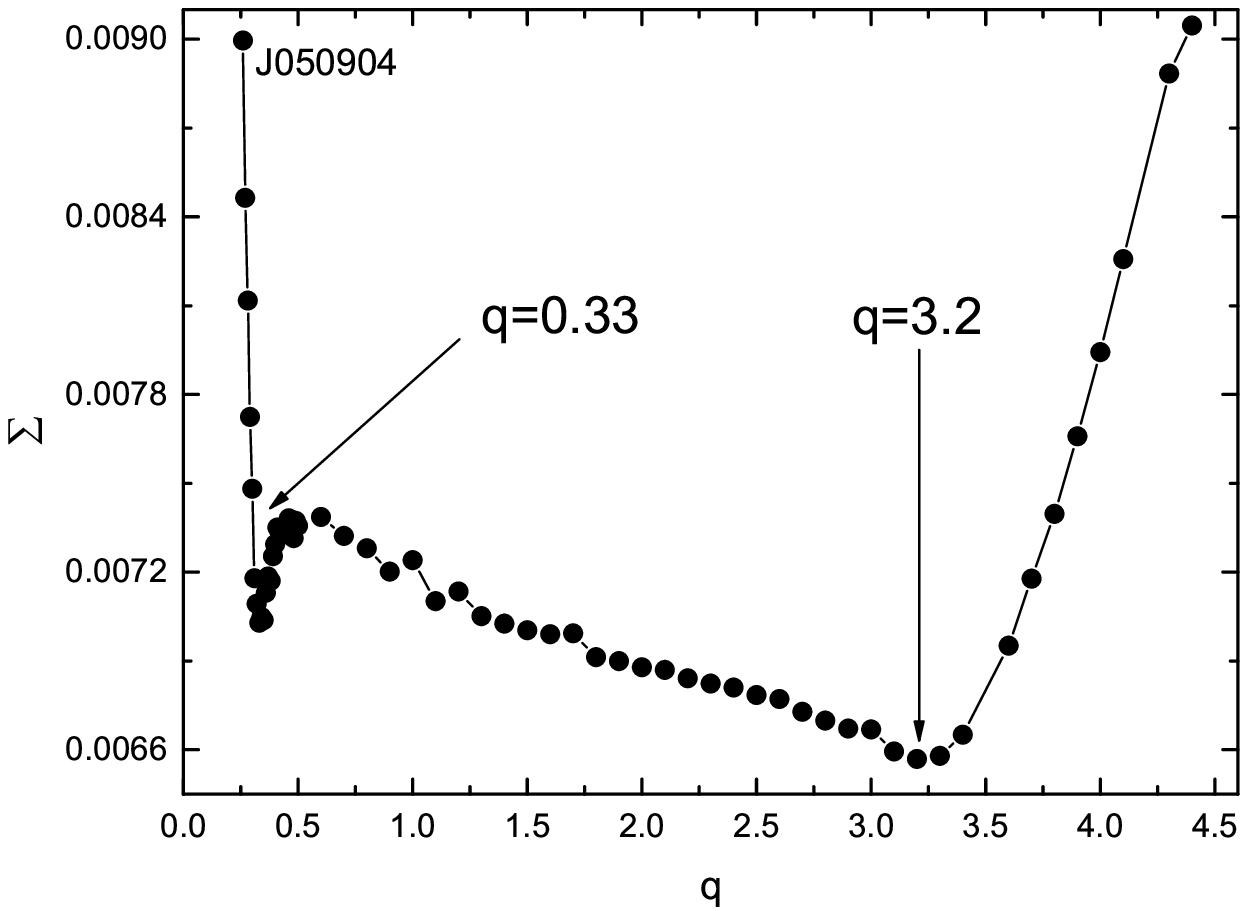}
\includegraphics[width=0.35\textwidth]{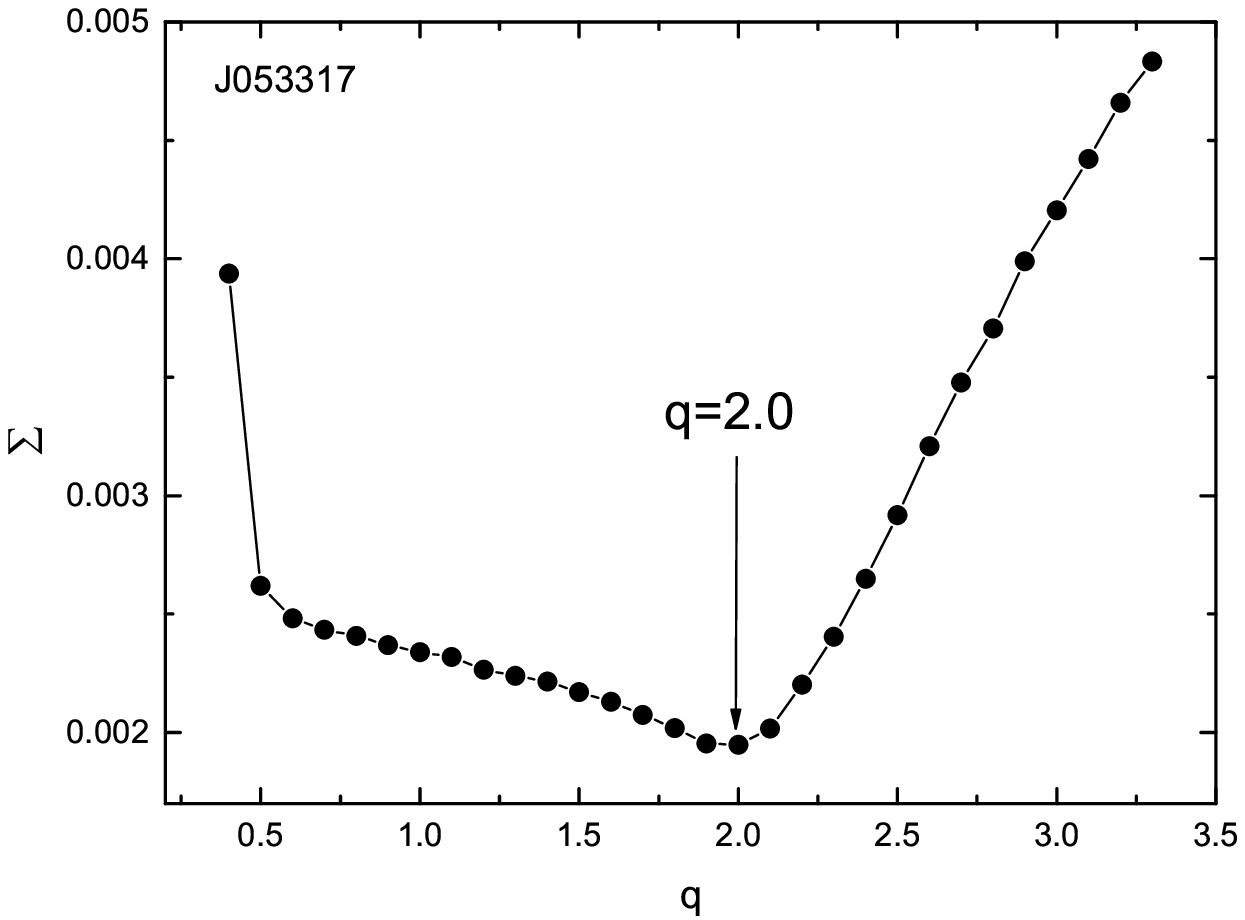}
\includegraphics[width=0.35\textwidth]{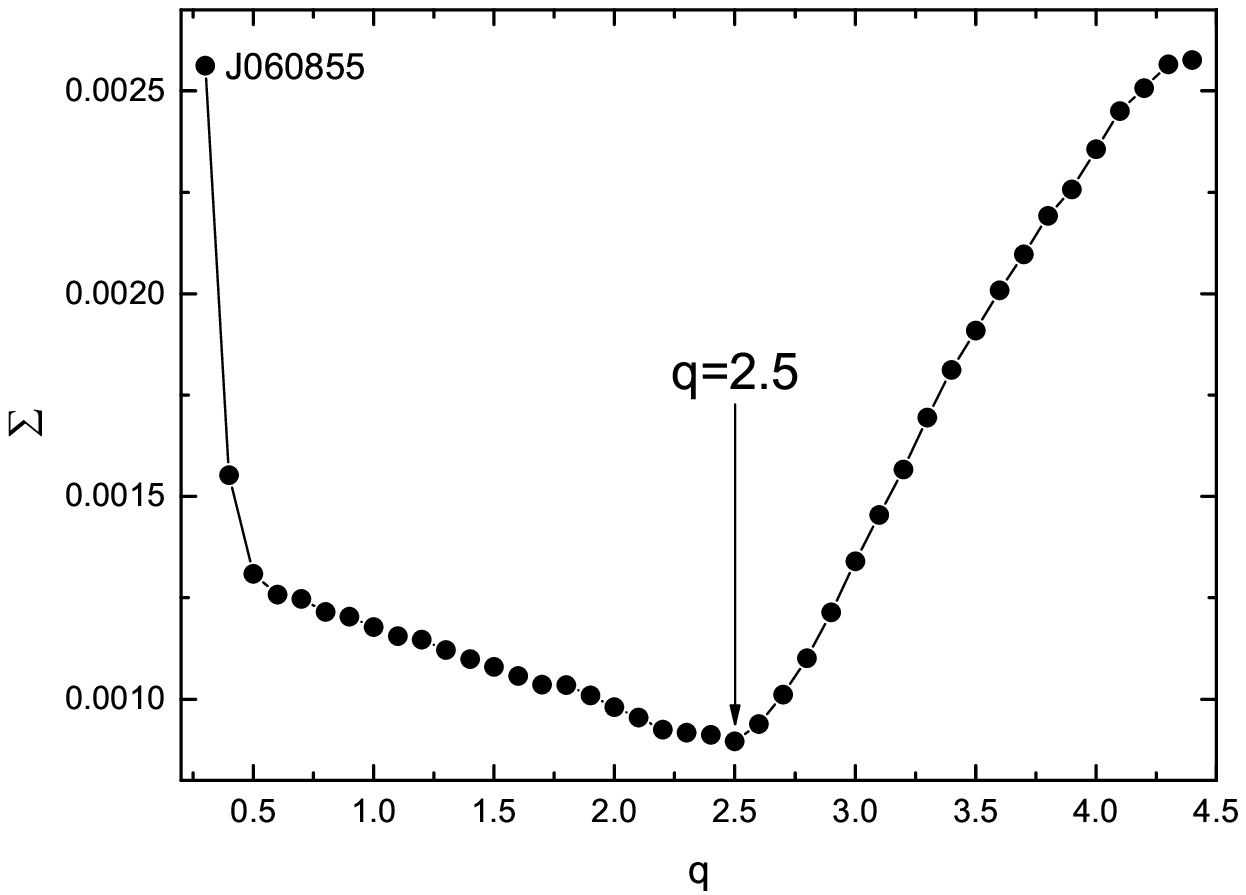}
\includegraphics[width=0.35\textwidth]{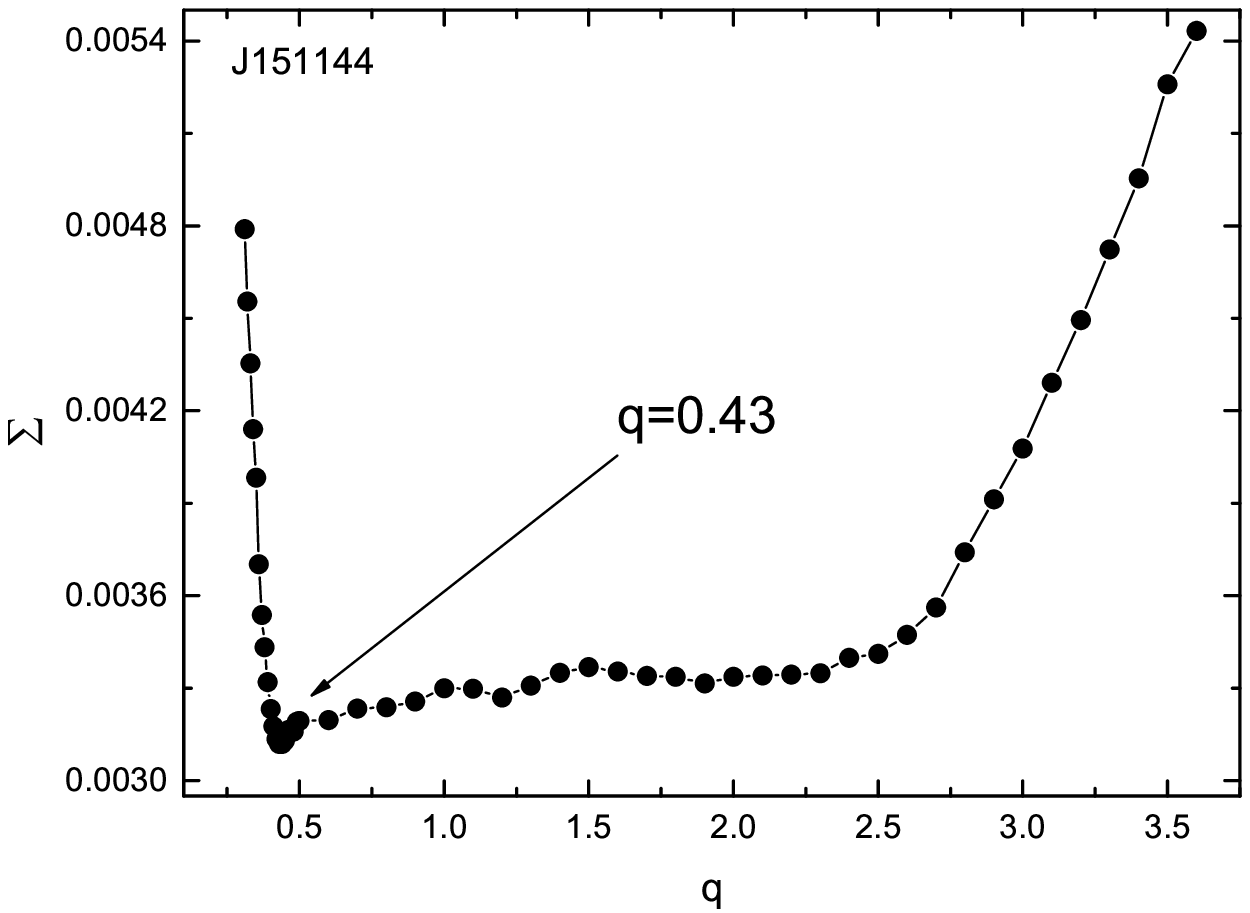}
\includegraphics[width=0.35\textwidth]{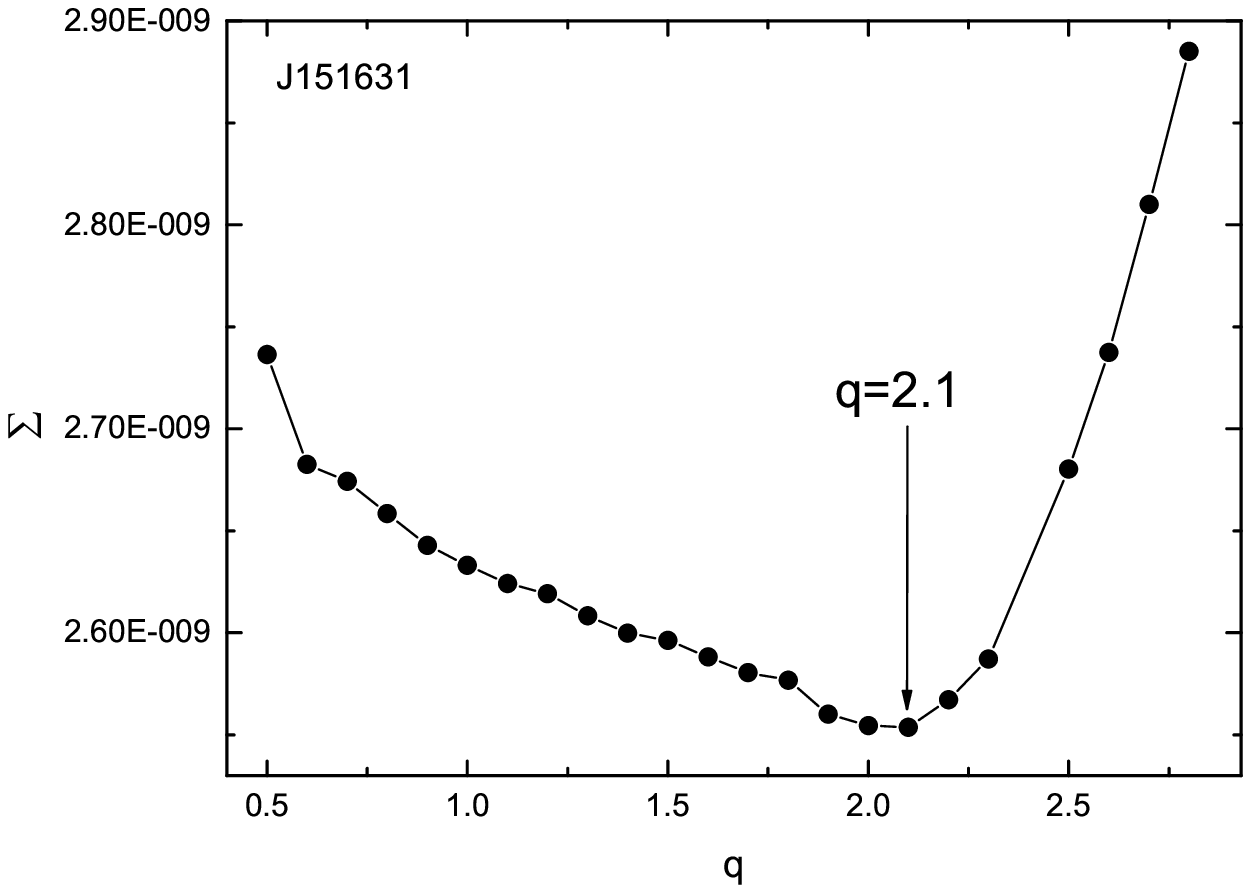}
\includegraphics[width=0.35\textwidth]{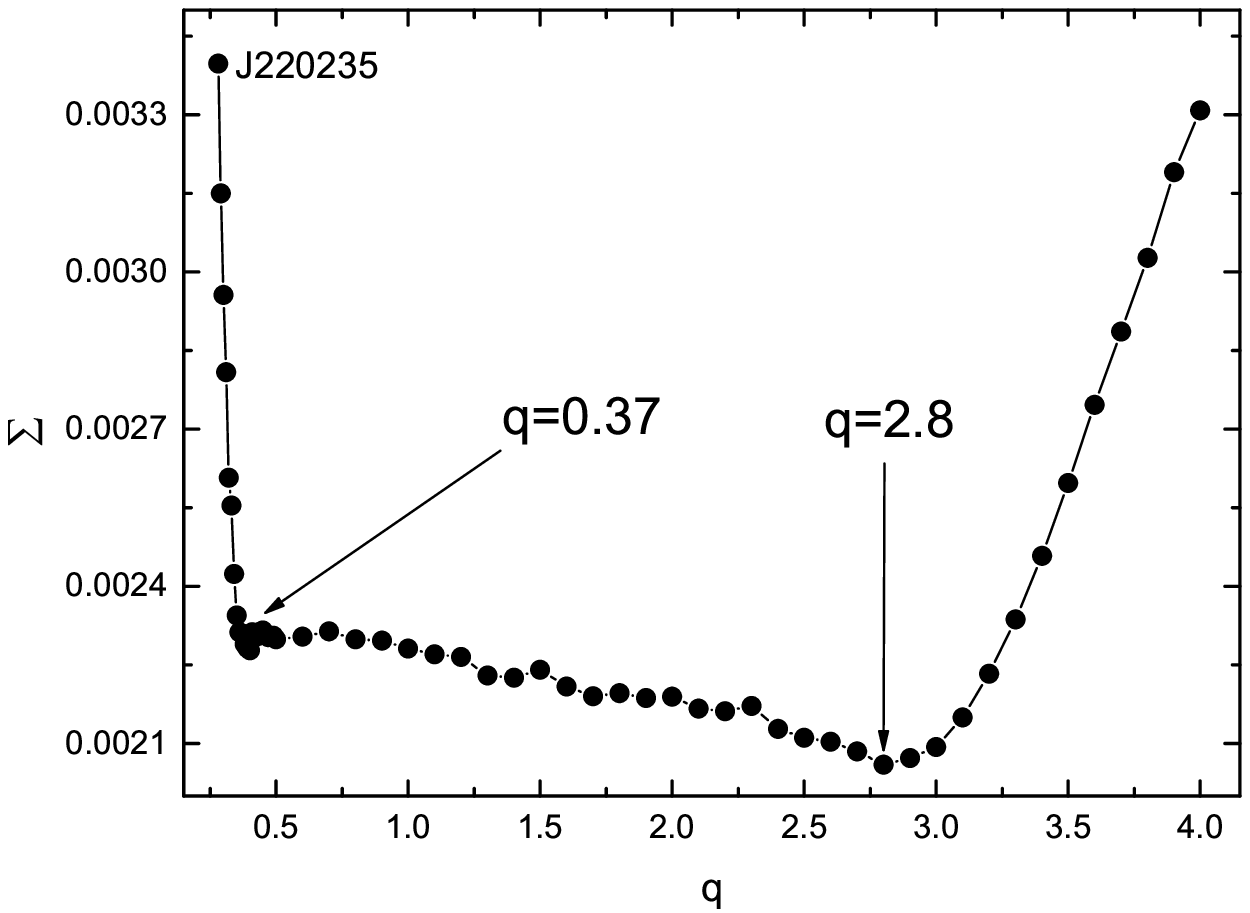}
\includegraphics[width=0.35\textwidth]{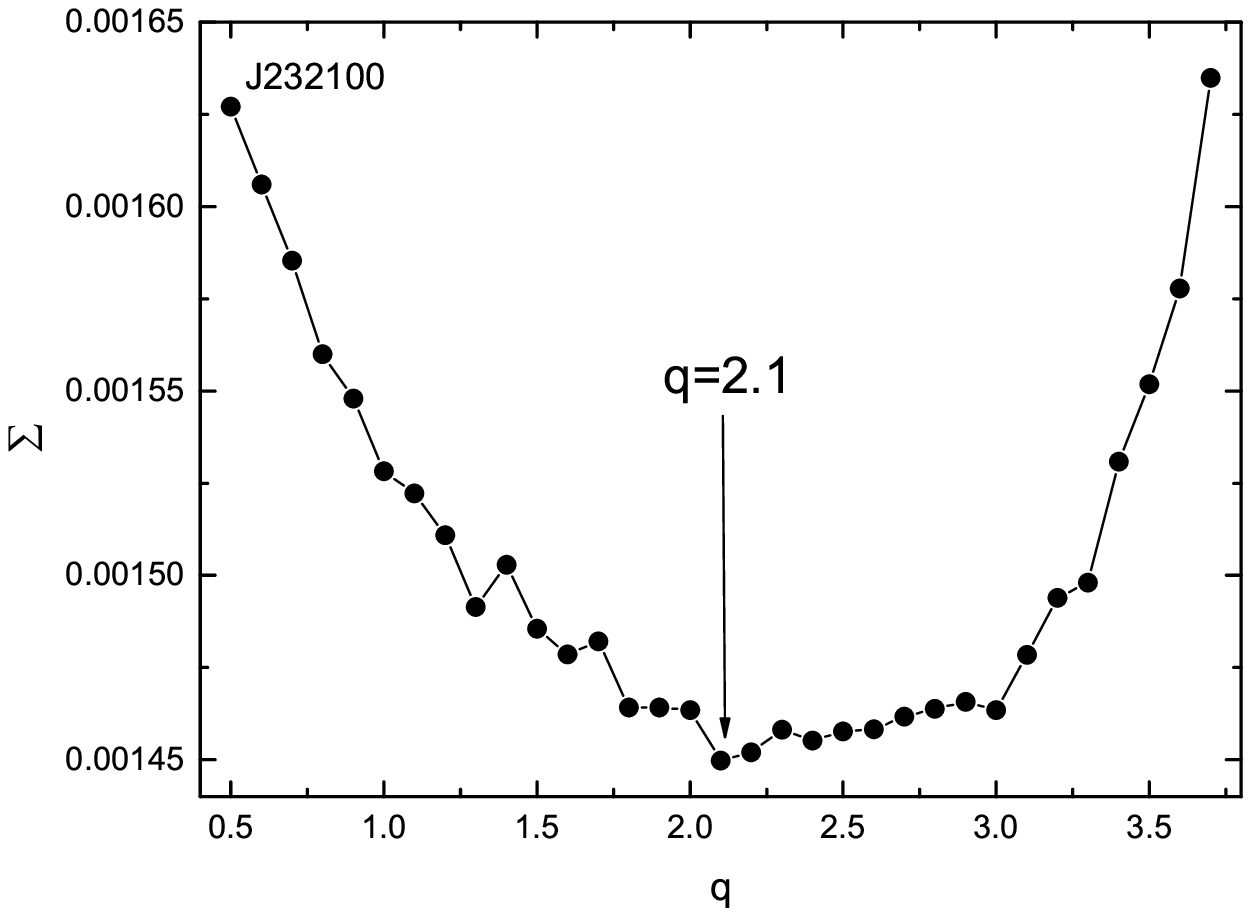}
\caption{ The relationship between resulting sum of the squares of the residuals and mass
ratio q for all the systems.}
\end{figure}

\begin{figure}[hpb]\centering
\includegraphics[width=0.45\textwidth]{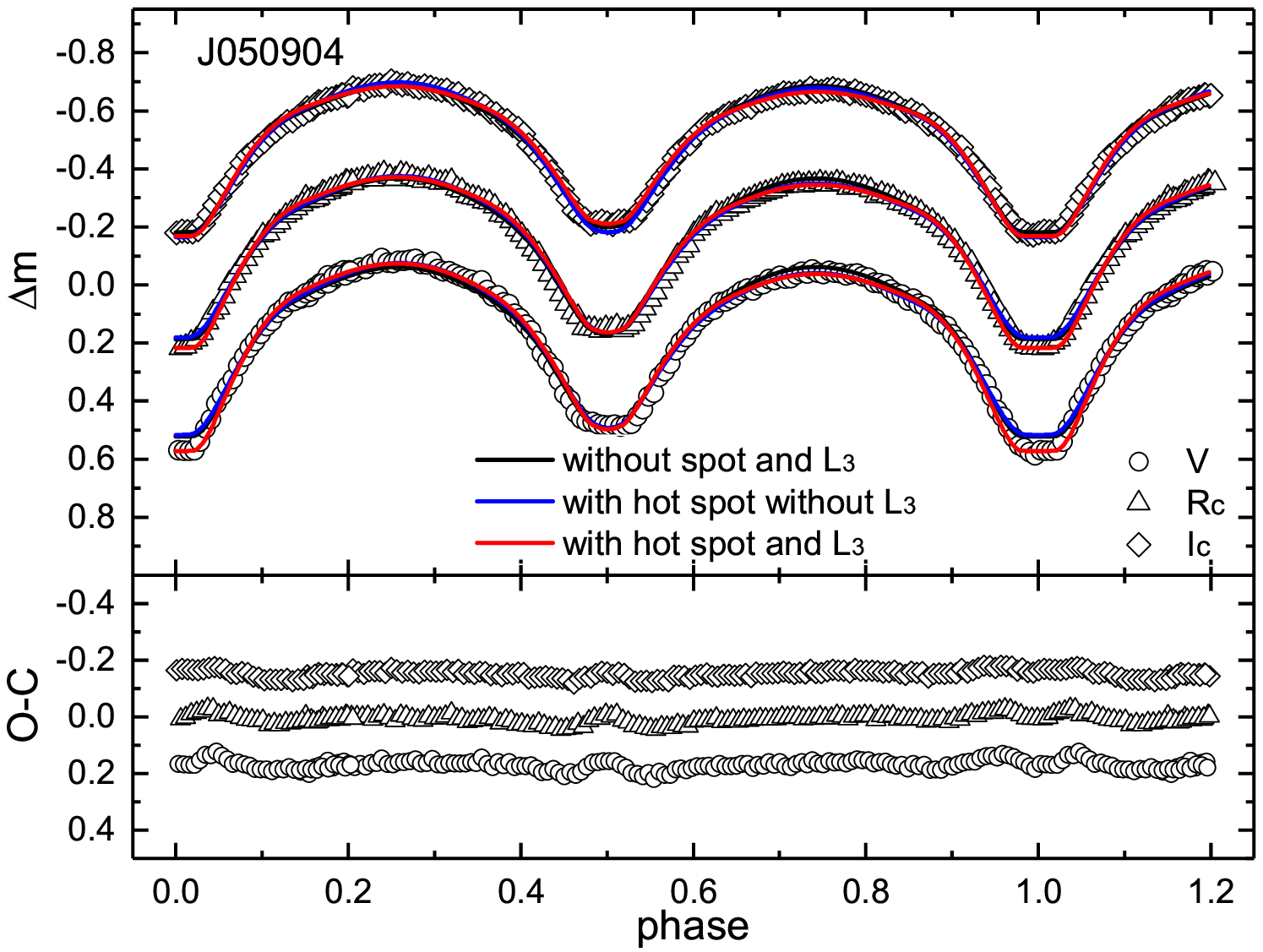}
\includegraphics[width=0.45\textwidth]{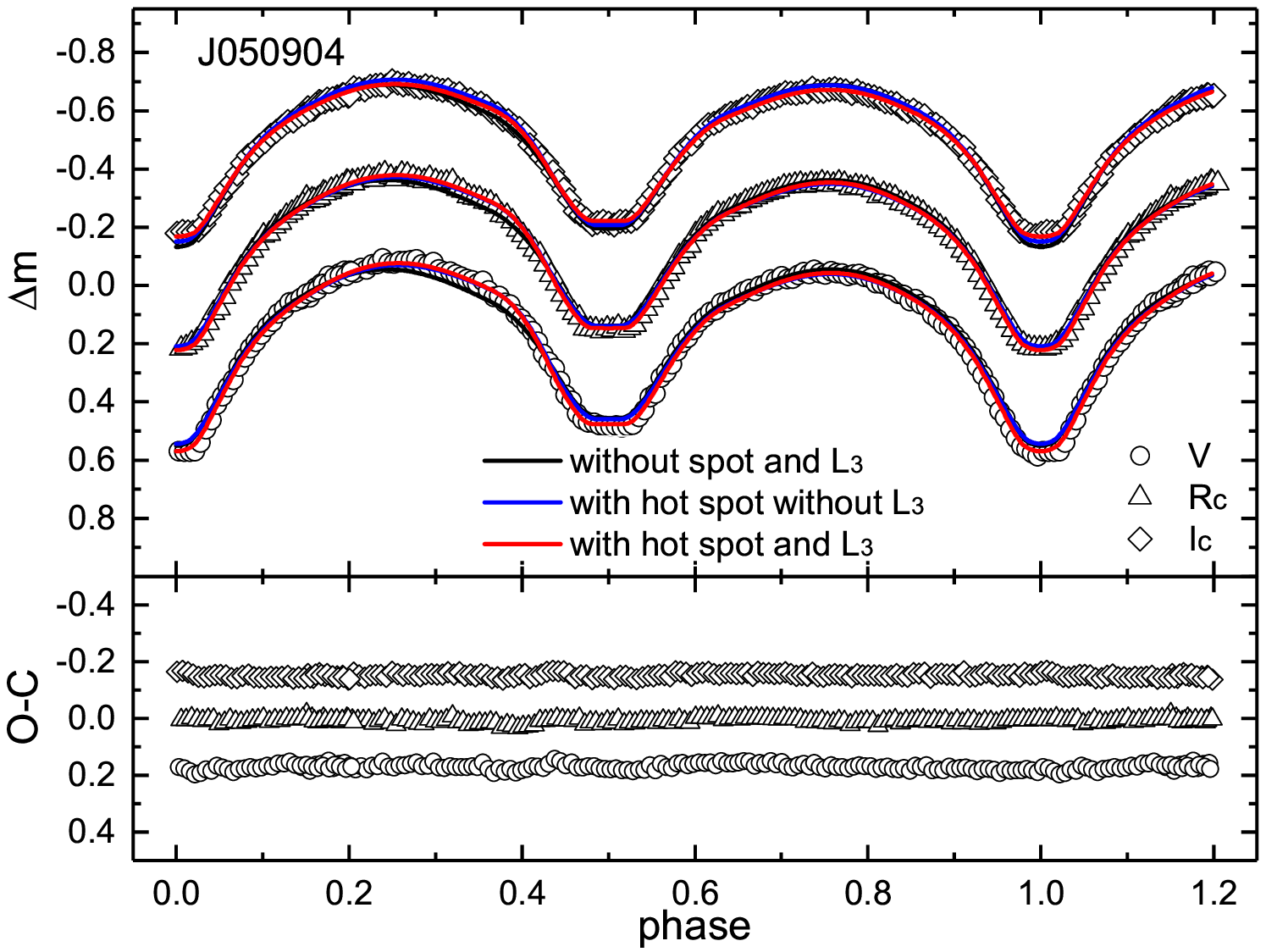}\\
\includegraphics[width=0.45\textwidth]{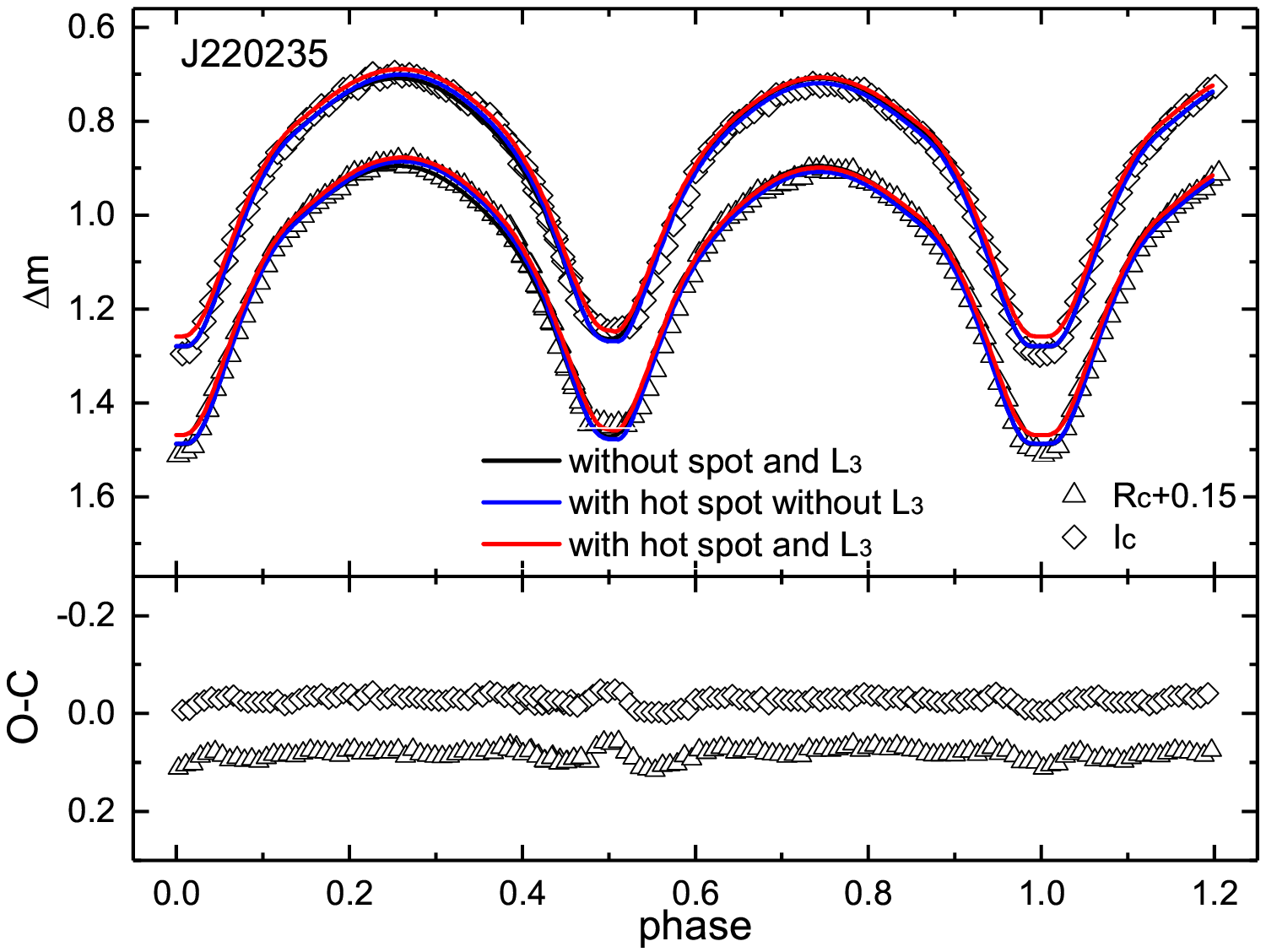}
\includegraphics[width=0.45\textwidth]{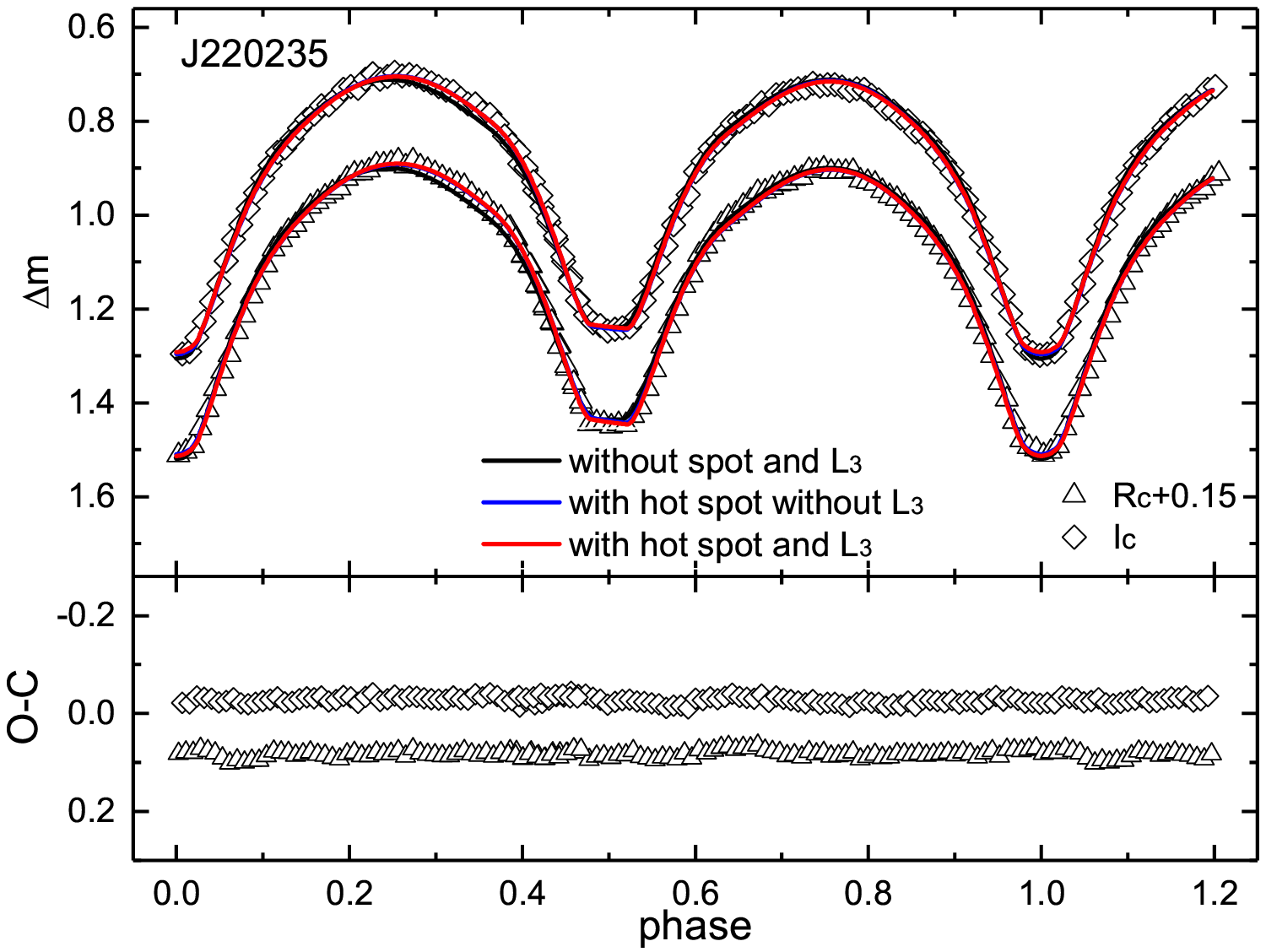}
\caption{Left panels show the observed and synthetic light curves determined by using the mass ratio of the minimum value of $\Sigma$ as an initial value, while right panels display the observed and synthetic light curves by using the mass ratio at the secondary minimum of the $q-$search figure as an initial value. The circles, triangles, and diamonds respectively refer to the $V$, $R_c$, and $I_c$ light curves, while the black, blue, and red lines respectively denote the synthetic light curves without spot and $L_3$, with spot and without $L_3$, and with spot and $L_3$. The phases of the two systems were
calculated using the period and HJD$_0$ listed in Table 1.}
\end{figure}

\begin{figure}[hpb]\centering
\includegraphics[width=0.40\textwidth]{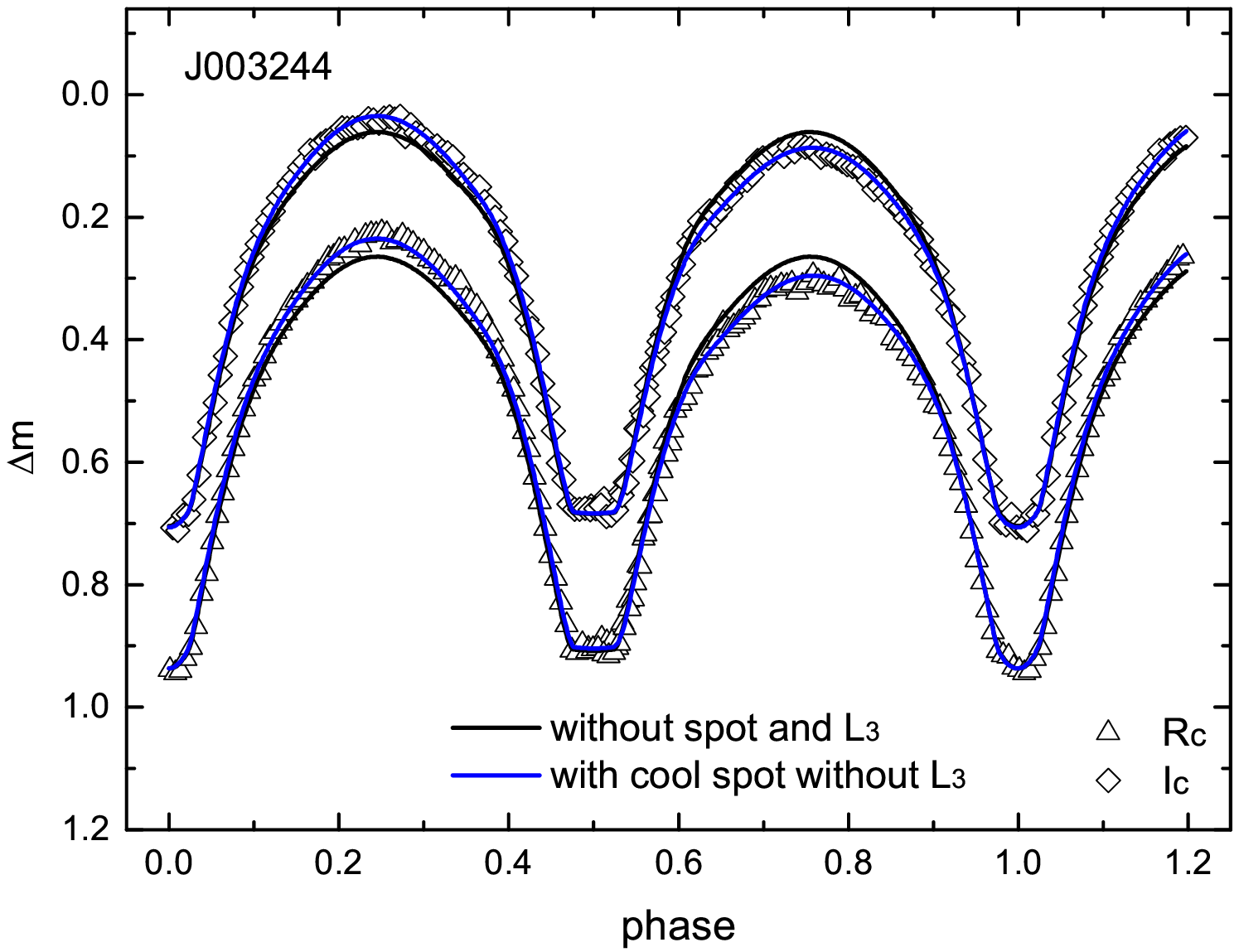}
\includegraphics[width=0.40\textwidth]{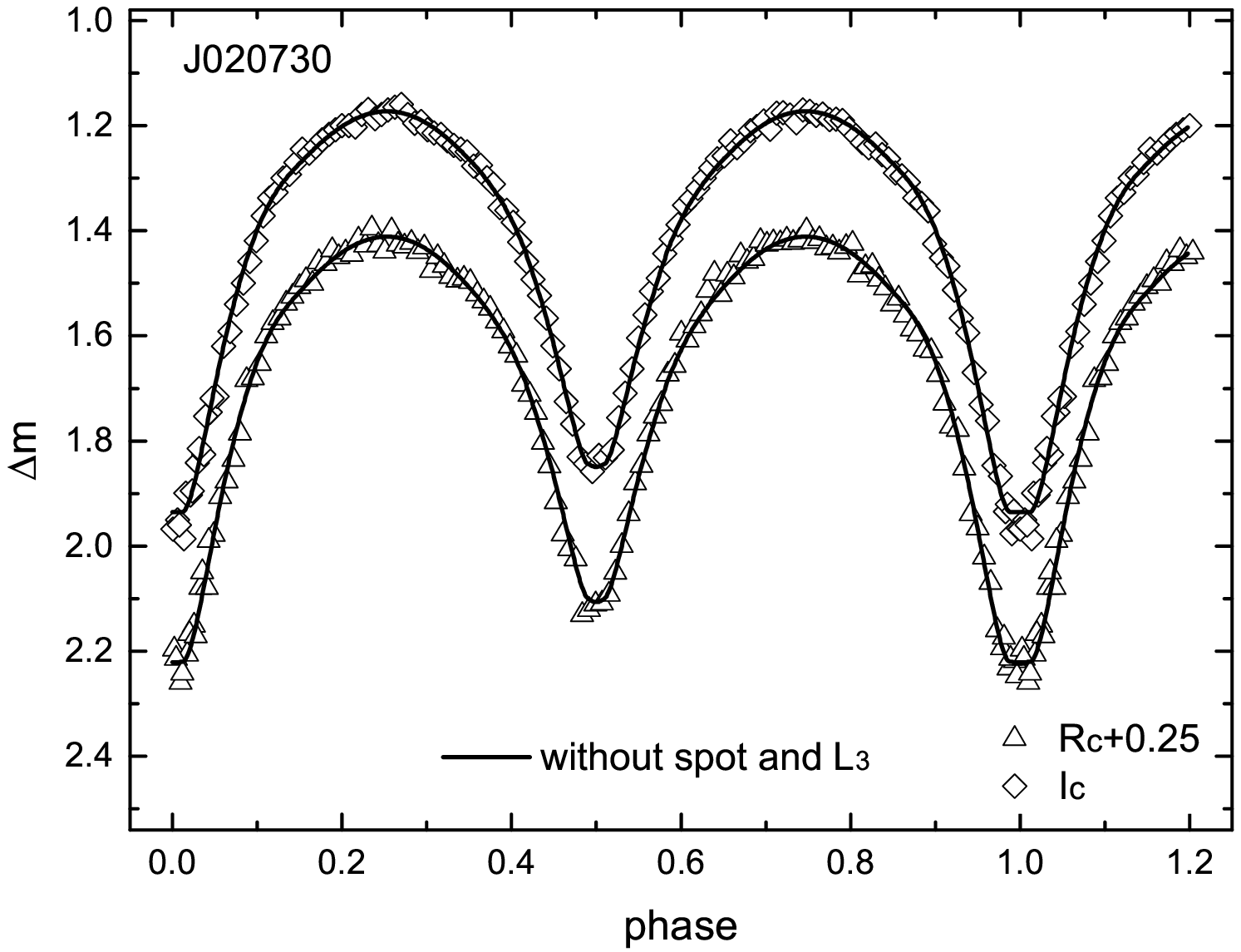}
\includegraphics[width=0.40\textwidth]{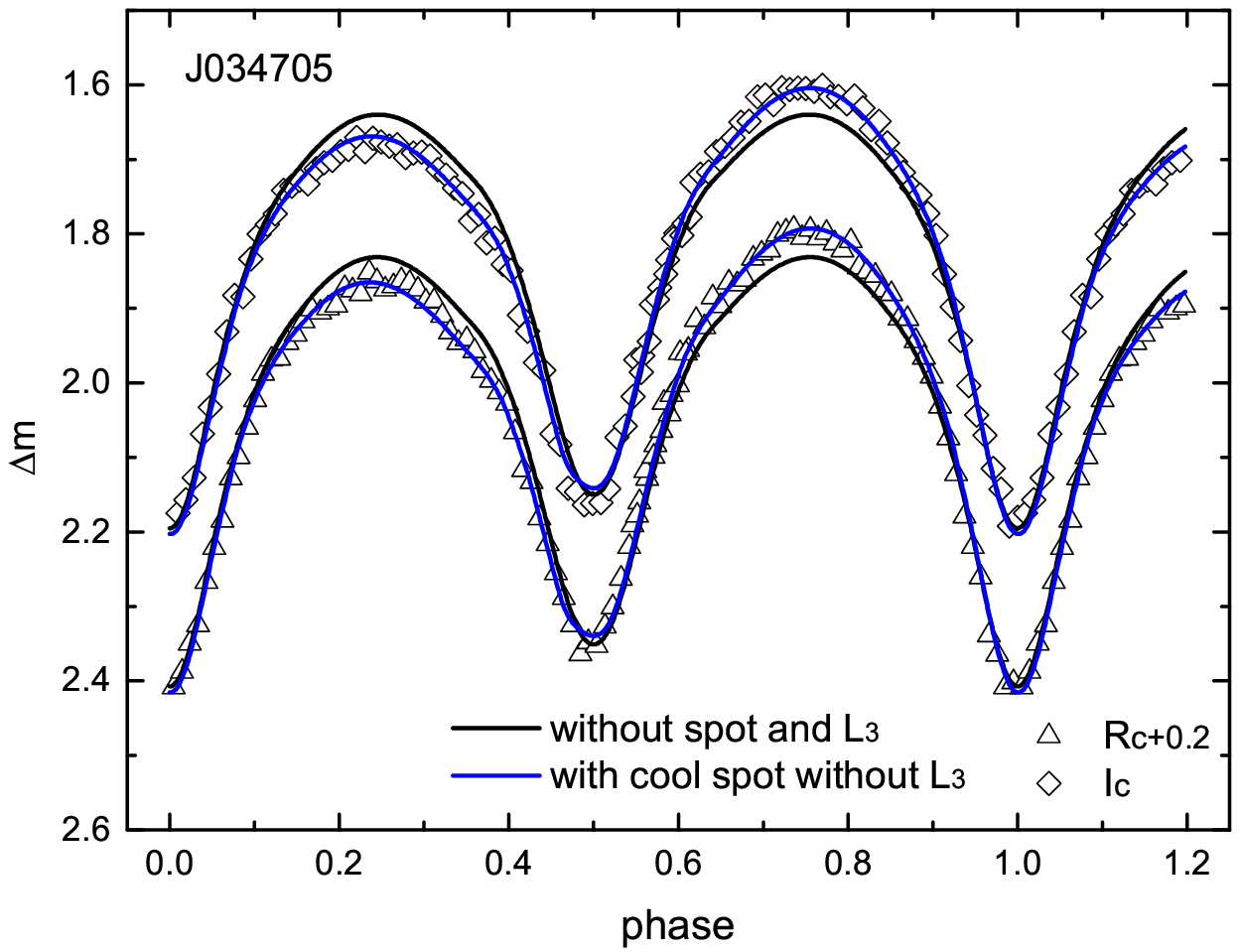}
\includegraphics[width=0.40\textwidth]{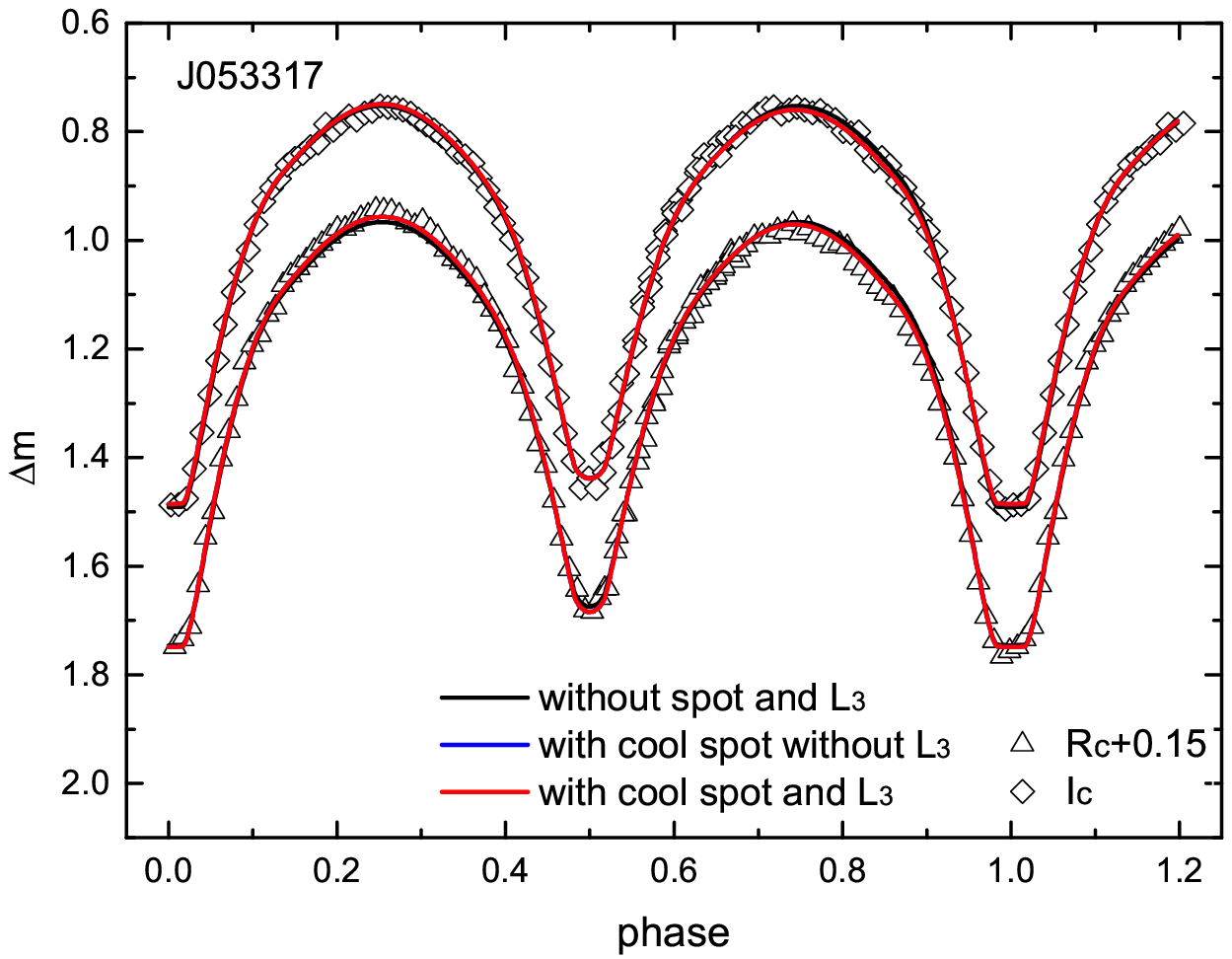}
\includegraphics[width=0.40\textwidth]{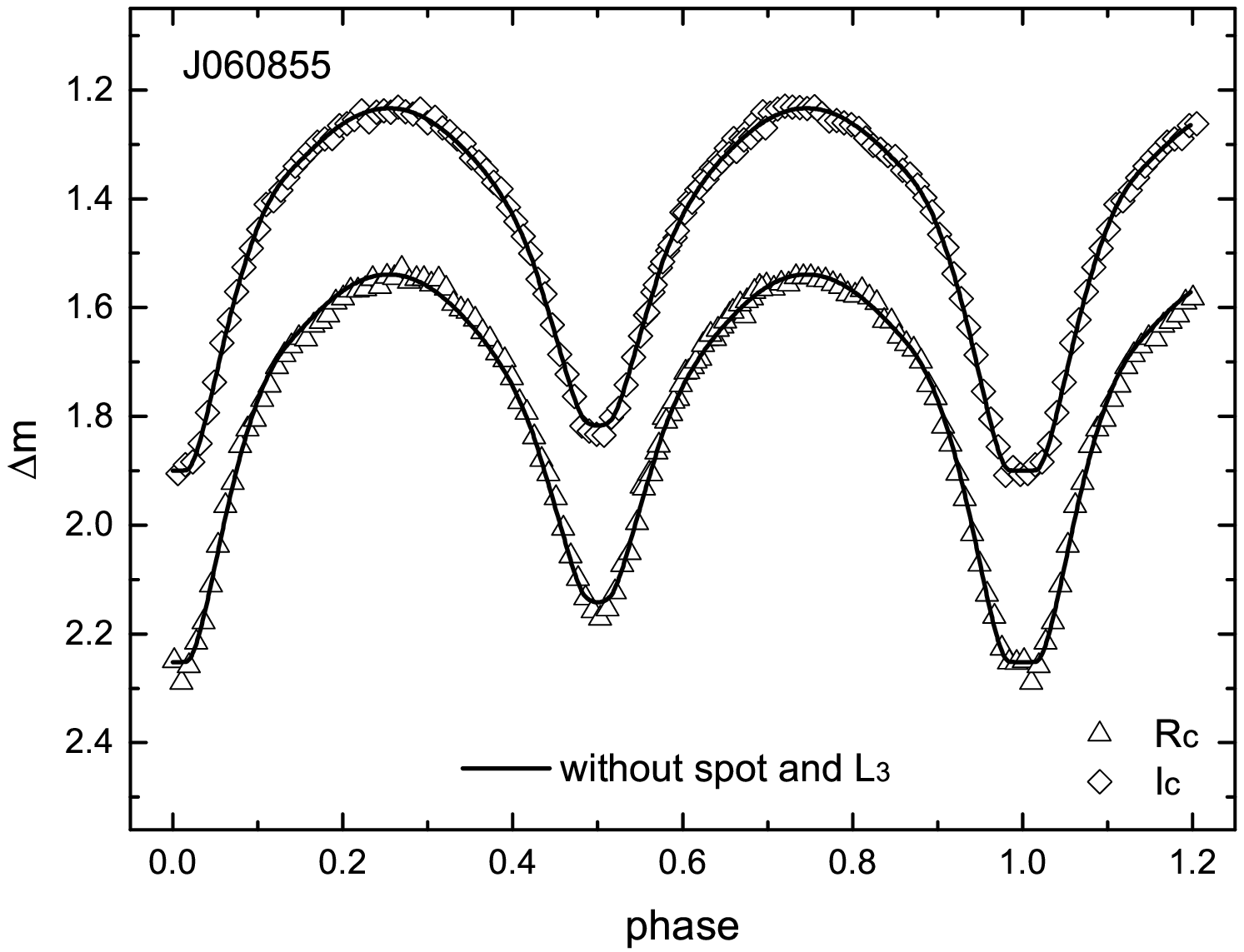}
\includegraphics[width=0.40\textwidth]{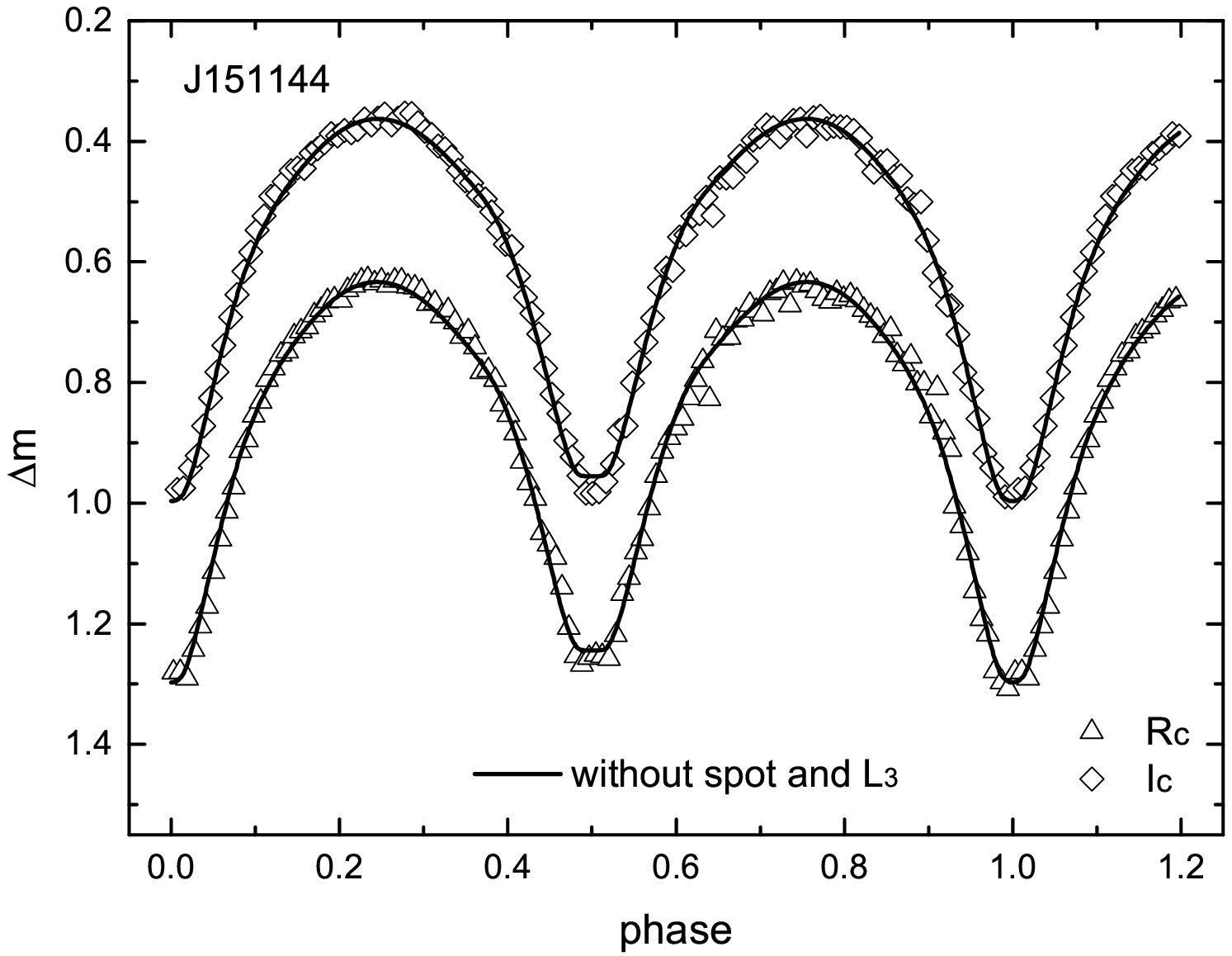}
\includegraphics[width=0.40\textwidth]{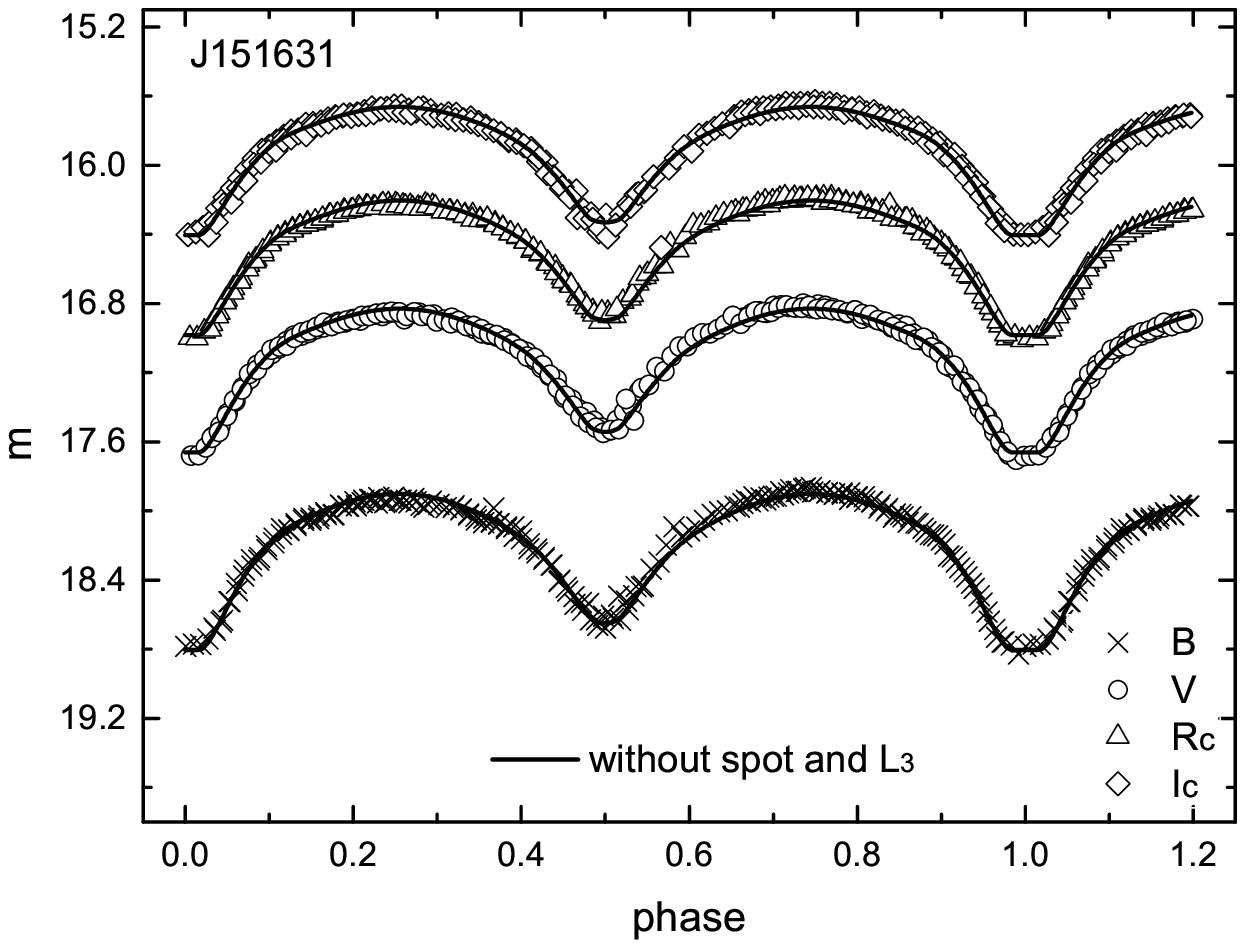}
\includegraphics[width=0.40\textwidth]{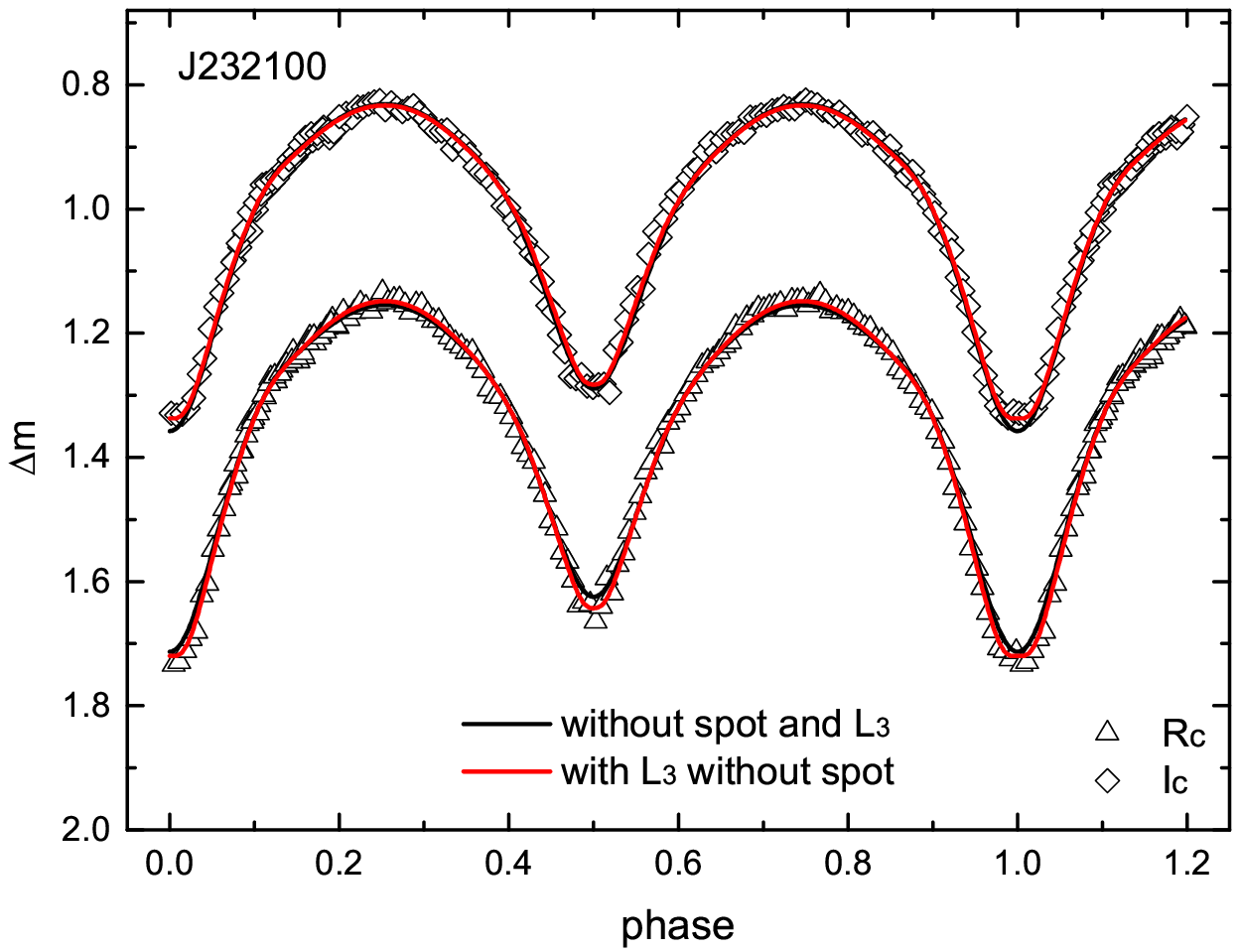}
\caption{The synthetic light curves for the other systems. The crosses, circles, triangles, and
diamonds respectively refer to the $B$, $V$, $R_c$, and $I_c$ light curves, while the black, blue, and
red lines respectively denote the synthetic light curves without spot and $L_3$, with spot and
without $L_3$, and with $L_3$ and with or without spot. The phases of all the systems were
calculated using the period and HJD$_0$ listed in Table 1.}
\end{figure}
\begin{table*}[!htb]
\tiny
\centering
\caption{Physical parameters of the targets}\label{tab:overfull}
\setlength{\leftskip}{-10pt}
\begin{tabular}{l|llllllllll}
\hline
\multicolumn{1}{c}{Star} & \multicolumn{ 2}{c}{J003244} &\multicolumn{ 1}{c}{J020730}  & \multicolumn{2}{c}{J034705}  &\multicolumn{3}{c}{J050904}  &  \multicolumn{ 2}{c}{J053317}              \\\hline
Mode                      & None        &  \multicolumn{1}{l|}{With S }   &  \multicolumn{1}{l|}{None }          &  None   &   \multicolumn{1}{l|}{With S }       &None    &  With S    &   \multicolumn{1}{l|}{With S, L$_3$  }    &  None   &  With S         \\\hline
$q(M_2/M_1) $        &  0.421(6)     &  \multicolumn{1}{l|}{0.443(3) }   &  \multicolumn{1}{l|}{1.928(26) }  &  0.405(23)   &   \multicolumn{1}{l|}{0.396(6) } &0.345(3)    &   0.330(3)    &   \multicolumn{1}{l|}{0.343(3) }    &   1.959(23)   &  1.883(13)         \\
$\Delta T(K)$         &  $-$87(15)      &  \multicolumn{1}{l|}{-80(6)	 }   &  \multicolumn{1}{l|}{175(6)    }  &  89(30)      &   \multicolumn{1}{l|}{94(12)   } &93(11)      &   83(8)       &   \multicolumn{1}{l|}{70(6)    }    &   141(7)      &  133(6)            \\
$i(deg)$             &  89.9(9)      &  \multicolumn{1}{l|}{89.9(1)  }   &  \multicolumn{1}{l|}{85.4(3)   }  &  78.5(9)     &   \multicolumn{1}{l|}{80.1(4)  } &81.4(3)     &   81.0(3)     &   \multicolumn{1}{l|}{82.7(3)  }    &   86.9(6)     &  87.8(3)           \\
$\Omega_{in} $       &  2.721        &  \multicolumn{1}{l|}{2.764 	 }   &  \multicolumn{1}{l|}{5.149     }  &  2.689       &   \multicolumn{1}{l|}{2.670    } &2.563       &   2.532       &   \multicolumn{1}{l|}{2.559    }    &   5.194       &  5.085             \\
$\Omega_{out}$       &  2.465        &  \multicolumn{1}{l|}{2.497 	 }   &  \multicolumn{1}{l|}{4.555     }  &  2.442       &   \multicolumn{1}{l|}{2.429    } &2.350       &   2.328       &   \multicolumn{1}{l|}{2.348    }    &   4.598       &  4.491             \\
$\Omega_1=\Omega_2$  &  2.676(15)    &  \multicolumn{1}{l|}{2.703(8) }   &  \multicolumn{1}{l|}{5.069(36) }  &  2.675(47)   &   \multicolumn{1}{l|}{2.633(11)} &2.535(7)    &   2.509(6)    &   \multicolumn{1}{l|}{2.528(6) }    &   5.111(32)   &  5.001(16)         \\
$L_{1B}/L_{B}$       &  -            &  \multicolumn{1}{l|}{-	       }   &  \multicolumn{1}{l|}{-         }  &  -           &   \multicolumn{1}{l|}{-        } &-           &   -           &   \multicolumn{1}{l|}{-        }    &   -           &  -                 \\
$L_{1V}/L_V$         &  -            &  \multicolumn{1}{l|}{-	       }   &  \multicolumn{1}{l|}{-         }  &  -           &   \multicolumn{1}{l|}{-        } &0.746(1)    &   0.749(1)    &   \multicolumn{1}{l|}{0.698(2) }    &   -           &  -                 \\
$L_{1R_c}/L_{R_c}$   &  0.667(1)     &  \multicolumn{1}{l|}{0.655(1) }   &  \multicolumn{1}{l|}{0.416(2)  }  &  0.718(4)    &   \multicolumn{1}{l|}{0.715(1) } &0.742(1)    &   0.746(1)    &   \multicolumn{1}{l|}{0.707(2) }    &   0.399(2)    &  0.406(1)          \\
$L_{1I_c}/L_{I_c}$   &  0.670(1)     &  \multicolumn{1}{l|}{0.658(1) }   &  \multicolumn{1}{l|}{0.403(2)  }  &  0.715(4)    &   \multicolumn{1}{l|}{0.712(1) } &0.740(1)    &   0.744(1)    &   \multicolumn{1}{l|}{0.709(2) }    &   0.388(2)    &  0.396(1)          \\
$L_{3B}/L_{B}$       &  -            &  \multicolumn{1}{l|}{-	       }   &  \multicolumn{1}{l|}{-         }  &  -           &   \multicolumn{1}{l|}{-        } &-           &   -           &   \multicolumn{1}{l|}{-        }    &   -           &  -                 \\
$L_{3V}/L_{V}$       &  -            &  \multicolumn{1}{l|}{-	       }   &  \multicolumn{1}{l|}{-         }  &  -           &   \multicolumn{1}{l|}{-        } &-           &   -           &   \multicolumn{1}{l|}{0.056(1) }    &   -           &  -                 \\
$L_{3R_c}/L_{R_c}$   &  -            &  \multicolumn{1}{l|}{-	       }   &  \multicolumn{1}{l|}{-         }  &  -           &   \multicolumn{1}{l|}{-        } &-           &   -           &   \multicolumn{1}{l|}{0.041(1) }    &   -           &  -                 \\
$L_{3I_c}/L_{I_c}$   &  -            &  \multicolumn{1}{l|}{-	       }   &  \multicolumn{1}{l|}{-         }  &  -           &   \multicolumn{1}{l|}{-        } &-           &   -           &   \multicolumn{1}{l|}{0.036(1) }    &   -           &  -                 \\
$r_1$                &  0.466(2)     &  \multicolumn{1}{l|}{0.465(1) }   &  \multicolumn{1}{l|}{0.331(1)  }  &  0.461(6)    &   \multicolumn{1}{l|}{0.471(1) } &0.482(1)    &   0.484(1)    &   \multicolumn{1}{l|}{0.482(1) }    &   0.330(1)    &  0.334(1)          \\
$r_2$                &  0.317(7)     &  \multicolumn{1}{l|}{0.325(4) }   &  \multicolumn{1}{l|}{0.447(6)  }  &  0.310(27)   &   \multicolumn{1}{l|}{0.310(8) } &0.296(4)    &   0.291(4)    &   \multicolumn{1}{l|}{0.301(5) }    &   0.447(5)    &  0.443(3)          \\
$f$                  &  17(6)\%      &  \multicolumn{1}{l|}{23(3)\%	 }   &  \multicolumn{1}{l|}{14(6)\%   }  &  6(19)\%     &   \multicolumn{1}{l|}{16(5)\%  } &13(3)\%     &   11(3)\%     &   \multicolumn{1}{l|}{15(3)\%  }    &   14(5)\%     &  14(3)\%           \\
Spot                 &               &  \multicolumn{1}{l|}{Star 2   }   &  \multicolumn{1}{l|}{          }  &              &   \multicolumn{1}{l|}{Star 2   } &            &   Star 2      &   \multicolumn{1}{l|}{Star 2   }    &               &  Star 1            \\
$\theta(deg)$        &  -            &  \multicolumn{1}{l|}{100(4) 	 }   &  \multicolumn{1}{l|}{-         }  &  -           &   \multicolumn{1}{l|}{67(6)    } &-           &   99(2)       &   \multicolumn{1}{l|}{99       }    &   -           &  47(4)             \\
$\lambda(deg)$       &  -            &  \multicolumn{1}{l|}{274(1) 	 }   &  \multicolumn{1}{l|}{-         }  &  -           &   \multicolumn{1}{l|}{80(2)    } &-           &   29(2)       &   \multicolumn{1}{l|}{29       }    &   -           &  40(7)             \\
$r_s(deg)$           &  -            &  \multicolumn{1}{l|}{25(1) 	 }   &  \multicolumn{1}{l|}{-         }  &  -           &   \multicolumn{1}{l|}{29(1)    } &-           &   17(1)       &   \multicolumn{1}{l|}{17       }    &   -           &  44(3)             \\
$T_s$                &  -            &  \multicolumn{1}{l|}{0.78(1)  }   &  \multicolumn{1}{l|}{-         }  &  -           &   \multicolumn{1}{l|}{0.72(2)  } &-           &   1.23(1)     &   \multicolumn{1}{l|}{1.23     }    &   -           &  0.98(1)           \\
$\Sigma W(O-C)^2$    &  0.0062       &  \multicolumn{1}{l|}{0.0024 	 }   &  \multicolumn{1}{l|}{0.0014    }  &  0.0020      &   \multicolumn{1}{l|}{0.0007   } &0.0066      &   0.0046      &   \multicolumn{1}{l|}{0.0040   }    &   0.0018      &  0.0015            \\
\hline
\multicolumn{1}{c}{Star}   & \multicolumn{1}{c}{J053317} & \multicolumn{1}{c}{J060855} &\multicolumn{1}{c}{J151144}  & \multicolumn{ 1}{c}{J151631}  &     \multicolumn{ 3}{c}{J220235}  & \multicolumn{ 2}{c}{J232100} & \\
\hline
Mode         & \multicolumn{1}{l|}{With S, L$_3$} & \multicolumn{1}{l|}{None }  &    \multicolumn{1}{l|}{None  }    &  \multicolumn{1}{l|}{None   }         &   None         &  With S  &   \multicolumn{1}{l|}{With S, L$_3$ }     &    None     & \multicolumn{1}{l|}{ With L$_3$}    \\\hline
$q(M_2/M_1) $        & \multicolumn{1}{l|}{1.883(28)} & \multicolumn{1}{l|}{2.486(27) }  &    \multicolumn{1}{l|}{0.409(7)  }    &  \multicolumn{1}{l|}{2.073(39)   }         &   0.377(4)         &  0.364(3)  &   \multicolumn{1}{l|}{0.381(6)  }     &    2.370(53)     & \multicolumn{1}{l|}{  2.237(58)}    \\
$\Delta T(K)$           & \multicolumn{1}{l|}{155(6)   } & \multicolumn{1}{l|}{191(6)    }  &    \multicolumn{1}{l|}{-2(11)    }    &  \multicolumn{1}{l|}{205(11)     }         &   79(12)           &  -102(7)   &   \multicolumn{1}{l|}{-108(6)   }     &    148(9)        & \multicolumn{1}{l|}{  175(6)   }    \\
$i(deg)$             & \multicolumn{1}{l|}{89.7(2)  } & \multicolumn{1}{l|}{83.3(3)   }  &    \multicolumn{1}{l|}{82.9(4)   }    &  \multicolumn{1}{l|}{85.9(7)     }         &   82.6(5)          &  83.2      &   \multicolumn{1}{l|}{84.5(5)   }     &    74.9(2)       & \multicolumn{1}{l|}{  82.8(6)  }    \\
$\Omega_{in} $       & \multicolumn{1}{l|}{5.085    } & \multicolumn{1}{l|}{5.926     }  &    \multicolumn{1}{l|}{2.698     }    &  \multicolumn{1}{l|}{5.354       }         &   2.630            &  2.604     &   \multicolumn{1}{l|}{2.638     }     &    5.767         & \multicolumn{1}{l|}{  5.583    }    \\
$\Omega_{out}$       & \multicolumn{1}{l|}{4.492    } & \multicolumn{1}{l|}{5.316     }  &    \multicolumn{1}{l|}{2.449     }    &  \multicolumn{1}{l|}{4.755       }         &   2.399            &  2.380     &   \multicolumn{1}{l|}{2.405     }     &    5.160         & \multicolumn{1}{l|}{  4.979    }    \\
$\Omega_1=\Omega_2$  & \multicolumn{1}{l|}{5.000(41)} & \multicolumn{1}{l|}{5.836(35) }  &    \multicolumn{1}{l|}{2.650(16) }    &  \multicolumn{1}{l|}{5.233(52)   }         &   2.597(8)         &  2.552(6)  &   \multicolumn{1}{l|}{2.584(9)  }     &    5.757(73)     & \multicolumn{1}{l|}{  5.490(86)}    \\
$L_{1B}/L_{B}$       & \multicolumn{1}{l|}{-        } & \multicolumn{1}{l|}{-         }  &    \multicolumn{1}{l|}{-         }    &  \multicolumn{1}{l|}{0.436(4)    }         &   -                &  -         &   \multicolumn{1}{l|}{-         }     &    -             & \multicolumn{1}{l|}{  -        }    \\
$L_{1V}/L_V$         & \multicolumn{1}{l|}{-        } & \multicolumn{1}{l|}{-         }  &    \multicolumn{1}{l|}{-         }    &  \multicolumn{1}{l|}{0.416(3)    }         &   -                &  -         &   \multicolumn{1}{l|}{-         }     &    -             & \multicolumn{1}{l|}{  -        }    \\
$L_{1R_c}/L_{R_c}$   & \multicolumn{1}{l|}{0.406(2) } & \multicolumn{1}{l|}{0.369(2)  }  &    \multicolumn{1}{l|}{0.688(1)  }    &  \multicolumn{1}{l|}{0.401(3)    }         &   0.724(1)         &  0.690(1)  &   \multicolumn{1}{l|}{0.668(4)  }     &    0.361(4)      & \multicolumn{1}{l|}{  0.310(2) }    \\
$L_{1I_c}/L_{I_c}$   & \multicolumn{1}{l|}{0.390(2) } & \multicolumn{1}{l|}{0.355(1)  }  &    \multicolumn{1}{l|}{0.688(1)  }    &  \multicolumn{1}{l|}{0.391(3)    }         &   0.722(1)         &  0.694(1)  &   \multicolumn{1}{l|}{0.663(4)  }     &    0.350(4)      & \multicolumn{1}{l|}{  0.283(2) }    \\
$L_{3B}/L_{B}$       & \multicolumn{1}{l|}{-        } & \multicolumn{1}{l|}{-         }  &    \multicolumn{1}{l|}{-         }    &  \multicolumn{1}{l|}{-           }         &   -                &  -         &   \multicolumn{1}{l|}{-         }     &    -             & \multicolumn{1}{l|}{  -        }    \\
$L_{3V}/L_{V}$       & \multicolumn{1}{l|}{-        } & \multicolumn{1}{l|}{-         }  &    \multicolumn{1}{l|}{-         }    &  \multicolumn{1}{l|}{-           }         &   -                &  -         &   \multicolumn{1}{l|}{-         }     &    -             & \multicolumn{1}{l|}{  -        }    \\
$L_{3R_c}/L_{R_c}$   & \multicolumn{1}{l|}{0.000(9) } & \multicolumn{1}{l|}{-         }  &    \multicolumn{1}{l|}{-         }    &  \multicolumn{1}{l|}{-           }         &   -                &  -         &   \multicolumn{1}{l|}{0.018(1)  }     &    -             & \multicolumn{1}{l|}{  0.181(1) }    \\
$L_{3I_c}/L_{I_c}$   & \multicolumn{1}{l|}{0.015(1) } & \multicolumn{1}{l|}{-         }  &    \multicolumn{1}{l|}{-         }    &  \multicolumn{1}{l|}{-           }         &   -                &  -         &   \multicolumn{1}{l|}{0.031(1)  }     &    -             & \multicolumn{1}{l|}{  0.225(1) }    \\
$r_1$                & \multicolumn{1}{l|}{0.334(1) } & \multicolumn{1}{l|}{0.311(1)  }  &    \multicolumn{1}{l|}{0.469(3)  }    &  \multicolumn{1}{l|}{0.330(2)    }         &   0.476(1)         &  0.483(1)  &   \multicolumn{1}{l|}{0.479(1)  }     &    0.306(2)      & \multicolumn{1}{l|}{  0.321(3) }    \\
$r_2$                & \multicolumn{1}{l|}{0.443(3) } & \multicolumn{1}{l|}{0.467(5)  }  &    \multicolumn{1}{l|}{0.314(9)  }    &  \multicolumn{1}{l|}{0.461(8)    }         &   0.300(5)         &  0.307(4)  &   \multicolumn{1}{l|}{0.311(7)  }     &    0.457(10)     & \multicolumn{1}{l|}{  0.457(12)}    \\
$f$                  & \multicolumn{1}{l|}{14(3)\%  } & \multicolumn{1}{l|}{15(6)\%   }  &    \multicolumn{1}{l|}{19(7)\%   }    &  \multicolumn{1}{l|}{20(9)\%     }         &   14(3)\%          &  23(3)\%   &   \multicolumn{1}{l|}{23(4)\%   }     &    2(12)\%       & \multicolumn{1}{l|}{  16(14)\% }    \\
Spot                 & \multicolumn{1}{l|}{Star 1   } & \multicolumn{1}{l|}{          }  &    \multicolumn{1}{l|}{          }    &  \multicolumn{1}{l|}{            }         &                    &  Star 1    &   \multicolumn{1}{l|}{Star 1    }     &                  & \multicolumn{1}{l|}{           }    \\
$\theta(deg)$        & \multicolumn{1}{l|}{47       } & \multicolumn{1}{l|}{-         }  &    \multicolumn{1}{l|}{-         }    &  \multicolumn{1}{l|}{-           }         &   -                &  133(1)    &   \multicolumn{1}{l|}{133       }     &    -             & \multicolumn{1}{l|}{  -        }    \\
$\lambda(deg)$       & \multicolumn{1}{l|}{40       } & \multicolumn{1}{l|}{-         }  &    \multicolumn{1}{l|}{-         }    &  \multicolumn{1}{l|}{-           }         &   -                &  204(1)    &   \multicolumn{1}{l|}{204       }     &    -             & \multicolumn{1}{l|}{  -        }    \\
$r_s(deg)$           & \multicolumn{1}{l|}{44       } & \multicolumn{1}{l|}{-         }  &    \multicolumn{1}{l|}{-         }    &  \multicolumn{1}{l|}{-           }         &   -                &  17(1)     &   \multicolumn{1}{l|}{17        }     &    -             & \multicolumn{1}{l|}{  -        }    \\
$T_s$                & \multicolumn{1}{l|}{0.98     } & \multicolumn{1}{l|}{-         }  &    \multicolumn{1}{l|}{-         }    &  \multicolumn{1}{l|}{-           }         &   -                &  1.22(1)   &   \multicolumn{1}{l|}{1.22      }     &    -             & \multicolumn{1}{l|}{  -        }    \\
$\Sigma W(O-C)^2$    & \multicolumn{1}{l|}{0.0013   } & \multicolumn{1}{l|}{0.0008    }  &    \multicolumn{1}{l|}{0.0029    }    &  \multicolumn{1}{l|}{0.0001      }         &   0.0021           &  0.0010    &   \multicolumn{1}{l|}{0.0009    }     &    0.0013        & \multicolumn{1}{l|}{  0.0010   }    \\
\hline
\end{tabular}

In this table, $q$ is the mass ratio and $q=M_2/M_1$, $\Delta T$ is the temperature difference between the two component and $\Delta T=T_1-T_2$, $i$ is the orbital inclination of the eclipsing system, $\Omega_{in}$ and $\Omega_{out}$ are the potentials of the inner and outer Lagrangian equipotential surfaces, $\Omega_1\,,\Omega_2$ are the actual potentials of the two components and $\Omega_1=\Omega_2$, $L_{1}/L$ is the luminosity ratio of the primary component, $L_{3}/L$ is the luminosity ratio of the third light, $r_1$ and $r_2$ are relative radii of the components, $f$ is the fill-out factor, $\theta(deg)$, $\lambda(deg)$, $r_s(deg)$, and $T_s$ are spot parameters.\\
The mode "None" represents light curve solutions with no spot or L$_3$, the mode "With S" refers to light curve solutions with spot and without L$_3$, the mode "With S, L$_3$" means light curve solutions with spot and L$_3$, while the mode "With L$_3$" shows light curve solutions with L$_3$.
\end{table*}

\begin{table*}
\centering
\scriptsize
\caption{The final temperatures of the two components of all the targets}
\begin{tabular}{lcccccccccc}
\hline
Star     & J003244 &J020730   & J034705  &J050904  &J053317  & J060855  &J151144  &J151631  & J220235  & J232100   \\\hline
$T_1$(K)  &4827(2) & 4414(4)  &4868(4)   & 5126(2) &4401(4)  &4387(4)   &4425(3)  &4946(7)  &4967(2)   &4431(4) \\
$T_2$(K)  &4907(8) & 4239(10) &4774(16)  & 5056(8) &4246(10) &4196(10)  &4427(14) &4741(18) &5075(8)   &4265(10)\\
\hline
\end{tabular}
\end{table*}

\section{Discussion and conclusions}
The light curves of ten short period contact binaries around the short period cut-off have been analyzed for the first time. We found that two systems are A-subtype contact binaries, and that eight systems are W-subtype contact binaries. A very interesting result of our investigation is that all the targets are shallow contact binaries (fill-out factor is less than 25\%). As listed in Table 4, $\Omega_{in}\,, \Omega_{out}$, and $\Omega$ describe the fill-out factor, and the fill-out factor of a contact binary can be calculated using the following equation, $f=(\Omega_{in}- \Omega)/( \Omega_{in}- \Omega_{out})$, where $\Omega_{in}$ is the potential of the inner Lagrangian equipotential surface, $\Omega_{out}$ is the potential of the outer Lagrangian equipotential surface, and $\Omega$ is the actual potential of the system. We also discovered that the temperature differences between the two components of all the ten targets are no more than 205 K, revealing that all these systems are in thermal contact. Five of the targets exhibit light curve asymmetries, indicating stellar spot activity. A hot or cool spot on one of the two components can well model the asymmetric light curves. In total, we found that four targets have evidence of a third light.

\subsection{The reliability of the photometric results}
As seen in Figures 4 and 5, the light curves of J003244, J034705, J050904, J151144, and J220235 show flat secondary minima, while those of J020730, J053317, J060855, J151631, and J232100 show flat primary minima, and the orbital inclinations of all objects are greater than 80$^\circ$ based on the final solutions of the light curves, these indicate that all the ten systems are totally eclipsing binaries. According to the analysis on contact binaries that are derived both spectroscopic and photometric mass ratios, Pribulla et al. (2003) found that the photometric mass ratios of totally eclipsing systems almost agree with their spectroscopic ones, while the photometric mass ratios of partly eclipsing systems deviate from their spectroscopic ones. Their work proposed that the physical parameters determined from photometric light curves only are very reliable and accurate for the systems have total eclipses. Terrell \& Wilson (2005) confirmed this result by their study.

In spite of this, Zhang et al. (2017) discussed the reliability of the light curve analysis and concluded that the degree of symmetry of the light curves and the sharpness of the bottom of the q-search figure will affect the reliability of the photometric results. They deduced that the light curve solutions are more reliable when the light curves are more symmetric and that the mass ratio is more reliable when the bottom of the q-search figure is sharper. As seen in Figure 3, the q-search curves of two targets (J034705 and J232100) show a nearly flat bottom, meaning that it is very difficult to provide a well-constrained value for q from photometry alone. In addition, the light curves of five systems (J003244, J034705, J050904, J053317, and J220235) are asymmetric, the photometric results are somewhat unreliable. For these six systems, it is difficult to determine reliable physical parameters from photometry alone. The overall system parameters obtained for these systems can be no more than an example of the possible fits for that system. There may well be similar fits obtainable with quite different parameters (including mass ratio) that differ significantly from the one quoted in each case. However, for J020730, J060855, J151144, and J151631, the light curves are on the whole symmetric and the bottom of each q-search figure is sharp, a certain minimum can be derived. Therefore, the light curve solutions of these four systems are generally reliable according to the two criteria proposed by Zhang et al. (2017).

\subsection{The statistics of USPCBs with periods less than 0.23 days}
At present, more and more USPCBs with periods less than 0.23 days have been studied. In order to comprehend the short period cut-off of contact binaries, we carried out a statistical work. Almost all well studied USPCBs with periods less than 0.23 days have been collected in Table 6 which shows the physical parameters of fifty-five USPCBs, including the period P, type (A means the more massive component is the hotter one, while W means the less massive component is the hotter ones), the mass ratio $q$ (smaller than 1.0), the orbital inclination $i$, the fill-out factor $f(\%)$ ($f(\%)$ relates to the fill-out factor of a contact binary and quantifies the degree of contact and describes the geometry configurations of contact binaries. For contact binaries, $f(\%)$ varies from 0 to 100\%), the effective temperatures of the two components $T_p$ and $T_s$ (where p means the more massive component, and s represent the less massive one), the relative radii $r_p$ and $r_s$, the mean densities of stars 1 and 2, the star spot activity. The mean densities were calculated by using the following equation taken from Mochnacki (1981),
\begin{equation}
\overline{\rho_1}={0.0189\over r_1^3(1+q)P^2}\,g\,cm^{-3}, \qquad \overline{\rho_2}={0.0189q\over r_2^3(1+q)P^2}\,g\,cm^{-3},
\end{equation}
where $q$ means the mass ratio and $P$ means the period. Based on our statistics, some common properties of USPCBs are found: (1) the components of all USPCBs are late spectral stars, from G1 to M4, (2) most of them show star spot activity, implying very strong surface activity, (3) most of the USPCBs are shallow or medium contact binaries, only a few of them are deep contact systems, (4) the mass ratios of all USPCBs are no less than 0.32, (5) the number of W-subtype systems is nearly equal to that of the A-subtype systems. Some points have been established by others, such as points 1 and 2 (e.g., Zhang et al. 2014), while point 5 is very different from the result determined before.

\begin{table*}
\tiny
\begin{center}
\caption{The properties of fifty-five well studied UPCBs}
\begin{tabular}{p{3.7cm}p{0.8cm}p{0.4cm}p{0.4cm}p{0.4cm}p{0.4cm}p{0.4cm}p{0.4cm}p{0.4cm}p{0.4cm}p{0.9cm}p{0.9cm}p{0.3cm}p{1.1cm}}
\hline\hline
Star	                      &P 	&Type	    &$q$      &$i$	     &$f(\%)$	  &$T_p$	&$T_s$	&$r_p$	&$r_s$& $\overline{\rho_1}$& $\overline{\rho_2}$  &Spot  &Ref.   \\
&days&&&&&&&&&g cm$^{-3}$&g cm$^{-3}$&&\\\hline
CSS J214633.8+120016             &0.179327	&A	      &0.450	       &55.0	   &26      &3600	  &3600	  &0.471	&0.329&3.88	&5.12&Y         &(1)       \\
SDSS J001641-000925              &0.198563	&A	      &0.620	       &53.3	   &22      &4342	  &3889	  &0.454	&0.423&2.42	&2.42&N         &(2), (3)  \\
SDSS J012119.10-001949.9         &0.205200	&A	      &0.500	       &83.9	   &19      &3840	  &3812	  &0.492	&0.331&3.20	&4.12&N         &(4)       \\
1SWASP J200059.78+054408.9       &0.205691	&W	      &0.436	       &75.3	   &58      &4528	  &4800	  &0.505	&0.352&2.61	&3.10&Y         &(5)        \\
1SWASP J075102.16+342405.3       &0.209172	&A	      &0.780	       &77.0	   &96      &3950	  &3876	  &0.456	&0.473&1.89	&1.79&Y         &(6)        \\
NSVS 7179685                     &0.209740	&W	      &0.470	       &85.5	   &19      &3979	  &4100	  &0.454	&0.326&3.08	&3.96&Y         &(7)        \\
1SWASP J022727.03+115641.7       &0.210950	&W	      &0.464	       &83.0	   &10      &3759	  &3800	  &0.404	&0.320&3.11	&4.11&Y         &(8)        \\
ASAS 071829-0336.7               &0.211259	&A	      &0.830	       &76.4	   &8    	  &4025	  &3967	  &0.454	&0.371&3.51	&3.76&N         &(9)        \\
CRTS J232100.1+410736            &0.211984	&W	      &0.447	       &82.8	   &16      &4265	  &4431	  &0.489	&0.321&3.05	&3.93&N         &(10)       \\
1SWASP J234401.81-212229.1       &0.213676	&A	      &0.422	       &79.4	   &deep?   &4400	  &4400	  &0.418	&0.306&3.11	&4.27&Y         &(11), (12) \\
NSVS 9747584                     &0.214393	&W	      &0.324	       &77.4	   &17      &3838	  &3947	  &0.459	&0.297&2.62	&3.84&N         &(9)        \\
1SWASP J015100.23-100524.2       &0.214500	&W	      &0.320	       &79.4	   &15      &4366	  &4500	  &0.464	&0.292&2.67	&4.00&Y         &(13)       \\
CRTS J053317.3+014049            &0.215652	&W	      &0.531	       &89.7	   &14      &4246	  &4401	  &0.416	&0.334&3.05	&3.78&Y         &(10)       \\
NSVS 8626028                     &0.217407	&A	      &0.805	       &65.9	   &21      &4318	  &4095	  &0.407	&0.380&3.03	&3.25&Y         &(7)        \\
1SWASP J080150.03+471433.8       &0.217513	&A	      &0.473	       &82.7	   &29      &4589	  &4581	  &0.417	&0.329&2.80	&3.60&N         &(14)       \\
NSVS 4761821                     &0.217513	&W	      &0.432	       &83.8	   &14      &4685	  &4696	  &0.447	&0.318&2.79	&3.75&N         &(7)        \\
NSVS 925605                      &0.217629	&A	      &0.678	       &57.2	   &70      &3813	  &3135	  &0.479	&0.408&2.23	&2.37&Y         &(7)        \\
GSC 1387-0475                    &0.217811	&A	      &0.474	       &49.9	   &76      &4690	  &4524	  &0.485	&0.386&1.66	&2.23&N         &(15), (16) \\
1SWASP J133105.91+121538.0       &0.218010	&A	      &0.828	       &77.6	   &25      &4977	  &4664	  &0.440	&0.383&3.02	&3.20&Y         &(17)       \\
NSVS 4484038                     &0.218493	&A	      &0.792	       &70.1	   &9       &5000	  &4897	  &0.454	&0.364&3.28	&3.63&Y         &(7), (18)  \\
1SWASP J024148.62+372848.3       &0.219751	&W	      &0.813	       &68.7	   &23      &3877	  &3900	  &0.474	&0.380&2.98	&3.21&Y         &(19)       \\
1SWASP J151144.56+165426.4       &0.219865	&W	      &0.409	       &82.9	   &19      &4425	  &4427	  &0.438	&0.314&2.69	&3.67&N         &(10)       \\
1SWASP J220235.74+311909.7       &0.220477	&W	      &0.381	       &84.5	   &23      &4967	  &5075	  &0.423	&0.311&2.56	&3.57&Y         &(10)       \\
CC Com                           &0.220686	&W	      &0.526	       &89.8	   &17      &4200	  &4300	  &0.426	&0.335&2.85	&3.56&Y         &(20),(21)  \\
NSVS 10632802                    &0.220721	&W	      &0.457	       &79.1	   &39      &4576	  &4986	  &0.475	&0.344&2.42	&2.99&Y         &(9)        \\
1SWASP J074658.62+224448.5        &0.220850	&W	      &0.365	       &81.7	   &51      &4543	  &4717	  &0.445	&0.327&2.20	&2.96&Y         &(9), (22)  \\
Kepler 12055255                  &0.220950	&W	      &0.964	       &39.8	   &13      &3701	  &3744	  &0.440	&0.377&3.27	&3.55&Y         &(23)       \\
NSVS 2175434                     &0.220951	&A	      &0.337	       &91.4	   &24      &4903	  &4898	  &0.409	&0.297&2.54	&3.72&Y         &(9), (24)  \\
1SWASP J213252.93-441822.6       &0.221235	&W	      &0.560	       &81.9	   &14      &4671	  &4700	  &0.421	&0.339&2.91	&3.57&Y         &(5)        \\
CSS J171410.0+445850             &0.221462	&A	      &0.519	       &74.1	   &25      &5150	  &5080	  &0.395	&0.344&2.71	&3.23&Y         &(1)        \\
CSS J224326.0+154532             &0.223228	&A	      &0.410	       &60.6	   &25      &4500	  &4500	  &0.421	&0.323&2.53	&3.27&N         &(1)        \\
CRTS J020730.1+145623            &0.223442	&W	      &0.519	       &85.4	   &14      &4239	  &4414	  &0.393	&0.331&2.79	&3.56&N         &(10)       \\
NSVS 8040227                     &0.223714	&A	      &0.522	       &67.2	   &16      &4674	  &4430	  &0.383	&0.336&2.74	&3.41&Y         &(9)        \\
1SWASP J212808.86+151622.0       &0.224842	&W	      &0.400	       &72.7	   &12      &4350	  &4800	  &0.445	&0.288&3.18	&4.47&Y         &(25)       \\
1SWASP J140533.33+114639.1       &0.225126	&W	      &0.646	       &68.6	   &8       &4523	  &4680	  &0.496	&0.346&2.99	&3.53&Y         &(26)       \\
CRTS J003244.2+244707            &0.225173  &W        &0.443         &89.9     &23      &4827   &4907   &0.457	&0.325&2.57	&3.33&Y         &(10)       \\
NSVS 5038135                     &0.225435	&W	      &0.338	       &72.0	   &21      &4553	  &4704	  &0.491	&0.304&2.35	&3.34&Y         &(9), (27)  \\
1SWASP J160156.04+202821.6       &0.226529  &A        &0.670         &79.5     &10      &4500   &4500   &0.443	&0.358&2.85	&3.22&Y         &(28)       \\
1SWASP J052926.88+461147.5       &0.226642	&A	      &0.412	       &89.9	   &24      &5077	  &5071	  &0.474	&0.321&2.43	&3.25&Y         &(24)       \\
1SWASP J030749.87-365201.7       &0.226671	&W	      &0.439	       &78.2	   &0       &4697	  &4750	  &0.546	&0.314&2.91	&3.63&Y         &(5)        \\
CRTS J151631.0+382626            &0.227192  &W        &0.482         &85.9     &20      &4741   &4946   &0.469	&0.330&2.52	&3.31&N         &(10)       \\
1SWASP J093010.78+533859.5B      &0.227714  &A        &0.397         &86.0     &17      &4700   &4700   &0.479	&0.313&2.44	&3.37&Y         &(29), (30) \\
07g-3-00820                      &0.227876	&W	      &0.796	       &69.8	   &5       &4967	  &5100	  &0.505	&0.359&3.07	&3.49&N         &(31)       \\
V1104 Her                        &0.227876	&W	      &0.623	       &83.4	   &15      &3902	  &4050	  &0.392	&0.350&2.76	&3.25&Y         &(32)       \\
1SWASP J044132.96=440613.7       &0.228155  &A        &0.638         &87.7     &24      &4003   &3858   &0.447	&0.362&2.60	&2.98&Y         &(24)       \\
NSVS 2700153                     &0.228456	&A	      &0.775	       &47.8	   &7       &4785	  &4689	  &0.449	&0.364&2.98	&3.28&N         &(7)        \\
1SWASP J210318.76+021002.2       &0.228590  &A        &0.877         &81.9     &34      &5850   &5778   &0.465	&0.398&2.59	&2.69&Y         &(33)       \\
Kepler 8108785                   &0.228840	&A	      &0.831	       &61.8	   &0       &4711	  &4246	  &0.491	&0.360&3.20	&3.51&Y         &(23)       \\
NSVS 2729229                     &0.228840  &A        &0.810         &49.6     &25      &3942   &3692   &0.461	&0.384&2.67	&2.85&Y         &(9)        \\
1SWASP J200503.05-343726.5       &0.228884	&W	      &0.934	       &73.8	   &9       &4270	  &4500	  &0.475	&0.379&3.07	&3.20&Y         &(34)       \\
Kepler 1572353                   &0.228900	&A	      &0.958	       &43.8	   &0       &4525	  &4277	  &0.404	&0.374&3.28	&3.37&Y         &(23)       \\
CRTS J060855.6+622713            &0.229320	&W	      &0.402	       &83.3	   &15      &4196	  &4387	  &0.433	&0.311&2.52	&3.43&N         &(10)       \\
NSVS 2607629                     &0.229370  &A        &0.619         &80.2     &28      &5390   &5168   &0.467	&0.362&2.52	&2.90&Y         &(9), (35)   \\
CRTS J034705.9+211309            &0.229570  &A        &0.396         &80.1     &16      &4868   &4774   &0.471	&0.310&2.46	&3.41&Y         &(10)       \\
1SWASP J050904.45-074144.4       &0.229575  &A        &0.343         &82.7     &15      &5126   &5056   &0.482	&0.301&2.38	&3.36&Y         &(10)       \\
 \hline
\end{tabular}
\end{center}
(1) Kjurkchieva et al. 2016, (2) Davenport  et al. 2013, (3) Qian et al. 2015a, (4) Jiang et al. 2015a, (5) Liu et al. 2018, (6) Jiang et al. 2015b, (7) Dimitrov \& Kjurkchieva 2015,
(8) Liu et al. 2015a, (9) Kjurkchieva et al. 2018a, (10) This paper, (11) Lohr et al. 2013, (12) Koen 2014, (13) Qian et al. 2015b, (14)  Darwish et al. 2017, (15) Rucinski \& Pribulla 2008,
(16) Yang et al. 2010 , (17) Elkhateeb et al. 2014b, (18)  Zhang et al. 2014, (19) Jiang et al. 2015d, (20) Pribulla et al. 2007, (21) K\"{o}se et al. 2011, (22) Jiang et al. 2015c,
(23) Kjurkchieva \& Dimitrov 2015, (24) Kjurkchieva et al. 2018b, (25) Djura\v{s}evi\'{c} et al. 2016, (26) Zhang et al. 2018, (27) Ula\c{s} et al. 2018, (28) Lohr et al. 2014, (29) Koo et al. 2014,
(30) Lohr et al. 2015, (31) Gao et al. 2017, (32) Liu, N. P. et al. 2015b, (33) Elkhateeb et al. 2014a, (34) Zhang et al. 2017, (35) G\"{u}rol \& Michel 2017.
\end{table*}

\subsection{The period-color diagram}
Rucinski (1992) suggested that the fully convective limit causes the short period cut-off of contact binaries, and the full-convection limit was determined to be at about $B-V\simeq1.5$, which corresponds to a temperature of 3550 K (Pecaut \& Mamajek 2013). As shown in Table 6, only for NSVS925605, the less masive component has a temperature below 3550K. This indicates that the fully convective limit possibly causes the short period limit. Then, we plotted the relations between the period and the color of the two components in Figure 6. For comparison, we included the low mass contact binaries (LMCBs) shown in Yakut \& Eggleton (2005). The color index $B-V$ for the two components of all systems including the USPCBs and LMCBs were calculated by interpolating the values of temperature into Table 5 of Pecaut \& Mamajek (2013). In Figure 6, left panel displays the period-color diagram of USPCBs, while the right panel plots the period-color diagram of LMCBs, the solid circles represent the more massive components, while the open ones refer to the less massive ones. The black solid line in both panels is the the short-period blue envelope (SPBE) taken from Rucinski (1998). In the left panel, the black dashed line denotes the full-convection limit suggested by Rucinski (1992).

Then, we used a quadratic term to fit the left panel by the least-squares method and determined the following two equations,
\begin{eqnarray}
(B-V)_p=-0.63(\pm6.42)+27.64(\pm60.8)\times P -88.70(\pm143.69)\times P^2,
\end{eqnarray}
for the more massive component, and
\begin{eqnarray}
(B-V)_s=0.57(\pm6.93)+16.90(\pm65.65)\times P-64.84(\pm155.13)\times P^2,
\end{eqnarray}
for the less massive component. The two quadric fits are labeled with red solid and dashed lines, respectively. Due to the two equations and the fully convective limit ($B-V\simeq1.5$) predicted by Rucinski (1992), we determined $P_{minp}=0.1709379$ days and $P_{mins}=0.1818291$ days. The average value of $P_{min}=0.1763835$ days can be adopted as the possible shortest period of contact binaries. In our fifty-five sample, none of the systems have a period below the predicted value. This may indicate that the predicted shortest period is somewhat reliable and that the fully convective limit predicted by Rucinski (1992) is one possible reason for the short period cut-off. However, we should state that though the predicted period is below the observed ones of our fifty-five sample, this can not reveal a satisfactory prediction, maybe some physical processes have not been taken into account. In addition, the errors for the two quadric fits are really too large.

As seen in Figure 6, the period-color relation of USPCBs is different from that of LMCBs. For the LMCBs, most of them are redder than the SPBE provided by Rucinski (1998), this is consistent with the result determined by Rucinski (1998). But for the USPCBs, more than a half are bluer than the SPBE. As Rucinski (1998) claimed, the USPCBs may be metal poor stars. Therefore, the USPCBs are metal poor old stars in the universe.

\begin{figure*}
\begin{center}
\includegraphics[width=0.517\textwidth]{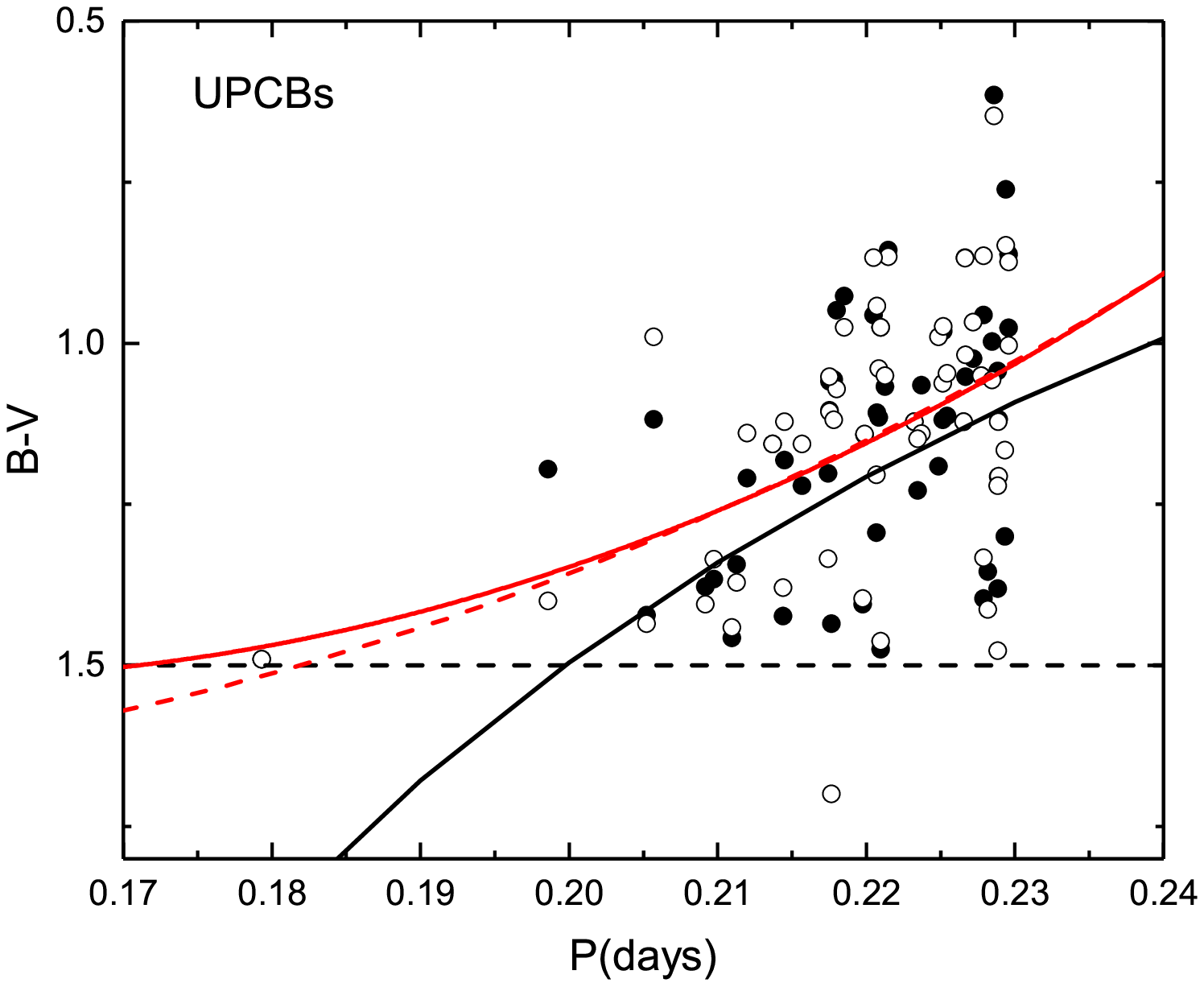}%
\includegraphics[width=0.5\textwidth]{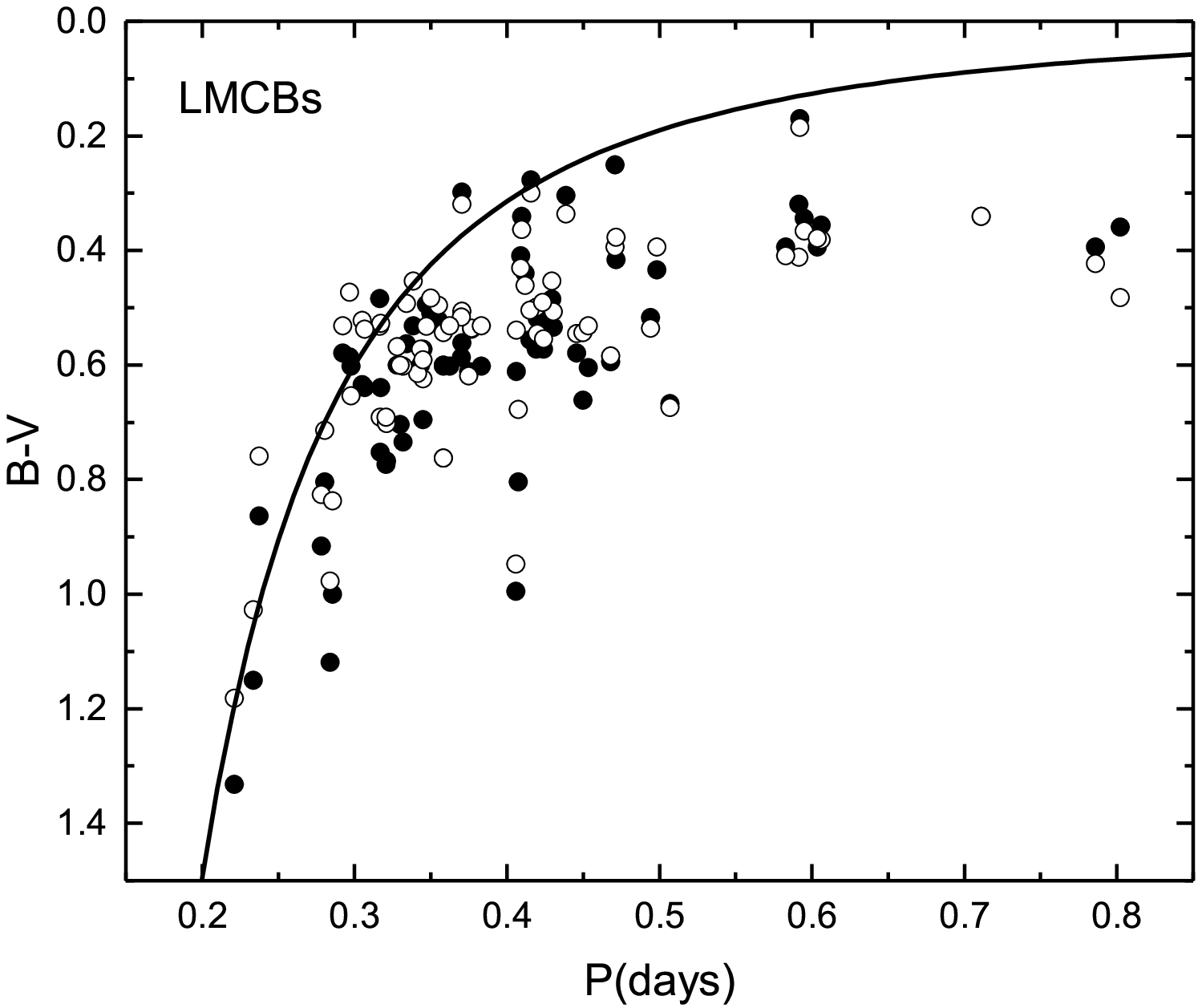}

\caption{Left panel displays the period-color diagram of USPCBs, while the right panel plots the period-color diagram of LMCBs. The solid circles represent the more massive components, while the open ones refer to the less massive ones. The black solid line in both panels is the SPBE taken from Rucinski (1998). In the left panel, the red solid and dashed lines reveal the quadratic fits to the more massive and less massive components, and black dashed line denotes the full-convection limit suggested by Rucinski (1992).}
\end{center}
\end{figure*}

\subsection{The evolution states of USPCBs}
All the USPCBs listed in Table 5 have no radial velocity curve observations, absolute parameters for them can not be directly derived. A direct method to study the evolutionary status of the USPCBs is the color-density diagram. We have determined the mean densities for both components as listed in Table 5, the color indices were also obtained. Then, the color-density diagram was constructed and displayed in Figure 7. The zero age main sequence (ZAMS) and the terminal age main sequence (TAMS) labeled with solid and dashed lines were extracted from Figure 3 of Mochnacki (1981). Red symbols represent the components of USPCBs, while black symbols show the components of LMCBs taken from Yakut \& Eggleton (2005). Not only for the LMCBs but also for the USPCBs, we found that the components of W-type contact binaries are generally more dense than that of A-type stars with the same spectral types. This agrees with the result determined by Mochnacki (1981) where A-type stars are less dense than ZAMS stars and W-type stars are usually closer to ZAMS stars. We also found that the less massive components of USPCBs and LMCBs lie upon the ZAMS line or between ZAMS and TAMS lines, which indicates that they are little-evolved or non-evolved stars. The more massive ones are located between the ZAMS and TAMS lines or below the TAMS line, which reveals that they are evolved TAMS or main sequence stars. The evolution states of USPCBs are identical to most W UMa contact binaries.
However, there are still some differences between USPCBs and LMCBs. Both the primary and secondary components of LMCBs are more evolved than those of USPCBs. This is caused by the masses of the components. The components of LMCBs should be more massive than those of the USPCBs. Therefore, they evolved faster than the less massive ones.

\begin{figure*}
\begin{center}
\includegraphics[angle=0,scale=0.7]{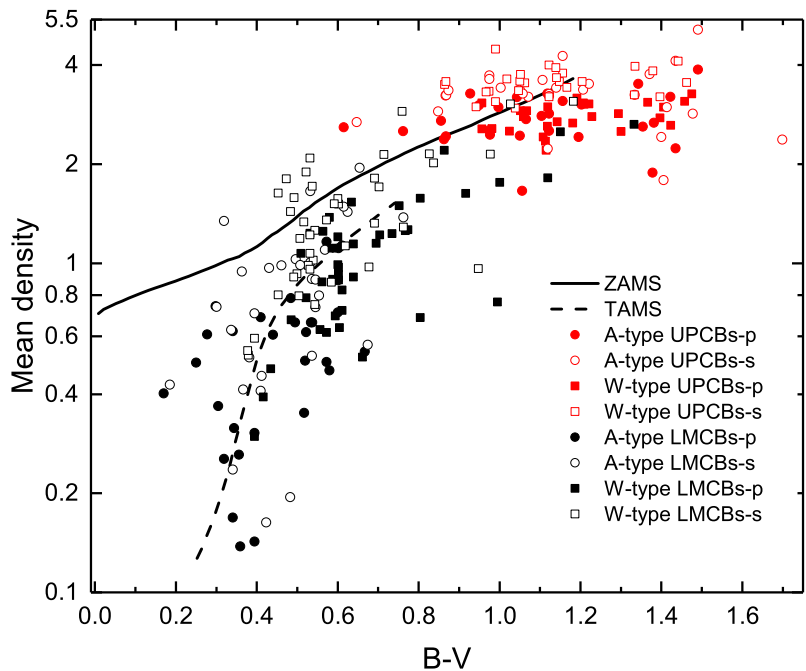}
\caption{This color-density diagram. Red solid circles and squares respectively show the more massive component of the A-type and W-type USPCBs, while the open circles and squares respectively display the less massive ones. Black solid circles and squares respectively show the more massive component of the A-type and W-type LMCBs taken from Yakut \& Eggleton (2005), while the open circles and squares respectively display the less massive ones. The ZAMS and TAMS lines extracted from Figure 3 of Mochnacki (1981) are respectively labeled with solid and dashed lines.
}
\end{center}
\end{figure*}

\subsection{Constraints to the explanation of the short period cut-off}
Based on the statistics, only one target has a temperature below the fully convective limit. The predicted shortest period of contact binaries due to the fully convective limit is below the observed ones of our fifty-five sample. These suggest that the fully convective limit proposed by Rucinski (1992) should be one reason for the short period cut-off.
Jiang et al. (2012) claimed that the mass limit of the primary component results in the short period cut-off. When the mass of the initial primary is less than 0.63 $M_\odot$, instability of mass transfer would occur when the primary fills its Roche lobe, and will result in quickly coalescence of both components. Because we can not directly determine the masses of the components, the explanation by Jiang et al. (2012) can not be accepted or rejected.

For the fifty-five well studied USPCBs, six are deep contact systems ($f>50\%$), forty-four are shallow contact systems ($f\leq25\%$). In addition, no system has a mass ratio of less than 0.32. This implies that most of them are at the beginning stage of contact, they are newly formed contact binaries. This is consistent with the AML theory proposed by Stepien (2006). He pointed out that the AML time scale is so long that low mass detached binaries can not fill their Roche lobes even at the universe time, and the total mass of the 0.22 d cut-off contact binaries is $\sim1.0-1.2$ $M_\odot$. According to the period-color relation of USPCBs, we found that the USPCBs are very ancient stars, this is in accordance with the very long AML time scale suggested by Stepien (2006).
However, the AML theory can not interpret the existence of the deep contact USPCBs. Other mechanism is required. Maybe a third body plays a very important role during the formation of the binary. In the six deep contact systems, five of them (1SWASP 074658.62+224448.5, 1SWASP J075102.16+342405.3, 1SWASP J234401.81-212229.1, NSVS 925605, GSC 1387-0475) are suggested to be in multiple systems. Statistic studies (e.g. Tokovinin et al. 2006; Pribulla \& Rucinski 2006; Liao et al. 2010) have shown that most contact binaries are members of multiple systems. During the formation of these six systems, third companions are removing angular momentum from their host eclipsing systems, this leads to very low angular momentum of the central eclipsing pair. Then, the initial short period detached eclipsing progenitors can evolve to deep contact systems by AML via magnetic stellar wind (Qian et al. 2007b, 2013, 2015a, 2017, 2018). A good way to search for a third body is via changes in the O-C diagram caused by light travel time effects. Three of the six systems (1SWASP 074658.62+224448.5, 1SWASP J075102.16+342405.3, 1SWASP J234401.81-212229.1) have been analyzed the O-C diagram, only one target, 1SWASP J234401.81-212229.1, has been detected cyclic variation (Lohr et al. 2013; Koen 2014). Though the other systems have not been detected cyclic variations in the O-C diagram, we cannot exclude this possibility because the time span of the observations of them is not long enough. We will continuously observe these six binaries in the future to search the cyclic variations in the O-C diagram.

In conclusion, based on the VSX, GCVS, CEV, and photometric surveys, we constructed the orbital period distribution of contact binaries, a sharp decline at 0.22 days is still present. By observing and analyzing ten totally eclipsing USPCBs, their photometric elements were obtained. A statistical work on well studied USPCBs was carried out, the common properties, the period-color relation, the evolution states were discussed. We suggested that both the fully convective limit claimed by Rucinski (1992) and the AML model proposed by Stepien (2006) can explain the short period limit, and for some cases, an additional companion plays a very important role. Because we collected only fifty-five USPCBs, the selection effect should not be neglected. Therefore, more observations and investigations of USPCBs are essential.

\acknowledgments
This work is supported by Chinese Natural Science Foundation (No. 11703016), and by the Joint Research Fund in Astronomy (No. U1431105) under cooperative agreement between the National Natural Science Foundation of China (NSFC) and Chinese Academy of Sciences (CAS), and by the Natural Science Foundation of Shandong Province (Nos. ZR2014AQ019, ZR2017PA009, ZR2017PA010, JQ201702), and by Young Scholars Program of Shandong University, Weihai (Nos. 20820162003, 20820171006), and by the Open Research Program of Key Laboratory for the Structure and Evolution of Celestial Objects (No. OP201704). RM acknowledges the financial support of project UNAM PAPIIT IN100918. Many thanks to the anonymous referee for very helpful comments and suggestions that improve our manuscript a lot.

We acknowledge the support of the staff of the Xinglong 85cm
telescope, SPMO 2.12m telescope, and WHOT. This work was partially supported by the Open Project Program of the Key
Laboratory of Optical Astronomy, National Astronomical Observatories, Chinese
Academy of Sciences.

The spectral data were provided by Guoshoujing Telescope (the Large Sky Area Multi-Object Fiber Spectroscopic Telescope LAMOST), which is a National Major Scientific Project built by the Chinese Academy of Sciences. Funding for the project has been provided by the National Development and Reform Commission. LAMOST is operated and managed by the National Astronomical Observatories, Chinese Academy of Sciences.

This work has made use of data from the European Space Agency (ESA) mission
{\it Gaia} (\url{https://www.cosmos.esa.int/gaia}), processed by the {\it Gaia}
Data Processing and Analysis Consortium (DPAC,
\url{https://www.cosmos.esa.int/web/gaia/dpac/consortium}). Funding for the DPAC
has been provided by national institutions, in particular the institutions
participating in the {\it Gaia} Multilateral Agreement.

\end{document}